\documentclass[12pt]{article}
\usepackage{enumerate}
\usepackage{amsfonts}
\usepackage{amsmath}
\usepackage{amssymb}
\usepackage{amsthm}

\def\H{\mathcal{H}}

\def\E{\mathcal{E}}
\def\P{\mathcal{P}}
\def\S{\mathfrak{S}}
\def\C{\mathfrak{C}}
\def\T{\mathfrak{T}}
\def\B{\mathfrak{B}}

\newcommand{\id}{\mathrm{Id}}
\newcommand{\Tr}{\mathrm{Tr}}

\newcommand{\shs}{\hspace{1pt}}

\newcounter{defin}  \newcounter{lemma} \newcounter{theorem}
\newcounter{property} \newcounter{corol}  \newcounter{remark} \newcounter{example}

\newenvironment{lemma}{\par\refstepcounter{lemma}
     \textbf{Lemma \thelemma.} }{\rm\par}

\newenvironment{property}{\par\refstepcounter{property}
     \textbf{Proposition \theproperty.}\ }{\rm\par}
\newenvironment{corollary}{\par\refstepcounter{corol}
     \textbf{Corollary \thecorol.} }{\rm\par}

\newenvironment{remark}{\par\refstepcounter{remark}
     \textbf{Remark \theremark.}}{\rm\par}

\begin{document}

\title{Tight continuity bounds for the quantum conditional mutual information, for the Holevo quantity and for capacities of quantum channels}
\author{M.E. Shirokov\footnote{Steklov Mathematical Institute, RAS, Moscow, email:msh@mi.ras.ru}}
\date{}
\maketitle

\begin{abstract}
We start with Fannes' type and Winter's type  tight continuity bounds for the quantum conditional mutual information and their specifications for states of special types.

Then we analyse continuity of the  Holevo quantity with respect to nonequivalent metrics on the set of discrete ensembles of quantum states. We show that the Holevo quantity is
continuous on the set of all ensembles of $m$ states with respect to all the metrics if either $m$ or the dimension of underlying Hilbert space is finite and obtain Fannes' type tight continuity bounds for the Holevo quantity in this case.

In general case conditions for local continuity of the Holevo quantity for discrete and continuous ensembles  are found. Winter's type  tight continuity bound for the Holevo quantity under  constraint on the average energy of ensembles is obtained and applied to the system  of quantum oscillators.

The above results are used  to obtain tight and close-to-tight continuity bounds for basic capacities of finite-dimensional channels (refining the Leung-Smith continuity bounds).
\end{abstract}

\pagebreak

\tableofcontents


\section{Introduction}

Quantitative analysis of continuity of basis  characteristics of quantum systems and channels is a necessary technical tool in study of their information properties. It suffices to mention that the famous Fannes continuity bound for the von Neumann entropy and the Alicki-Fannes continuity bound for the conditional entropy are essentially used in the proofs of several important results of quantum information theory \cite{H-SCI,N&Ch,Wilde}. During the last decade many papers devoted to finding continuity bounds (estimates for variation) for different  quantities have been appeared (see \cite{Aud,A&E,A&E+,L&S,W&R,W-CB} and the references therein).

Although in many applications a structure of  continuity bound of a given quantity is more important than concrete values of its coefficients, a task of finding  optimal values of these coefficients seems interesting from  both mathematical and physical points of view. This task can be formulated as a problem of finding so called "tight" continuity bound, i.e. relatively $\varepsilon$-sharp estimates for variations of a given quantity. The most known decision of this problem is the sharpest continuity bound for the von Neumann entropy obtained by Audenaert \cite{Aud} (it refines the Fannes continuity bound mentioned above). Other result in this direction is the tight bound for the relative entropy difference via the entropy difference obtained by Reeb and Wolf \cite{W&R}. Recently Winter presented tight continuity bound for the conditional entropy (improving the Alicki-Fannes continuity bound) and tight continuity bounds for the entropy and for the conditional entropy in infinite-dimensional systems under energy constraint \cite{W-CB}. By using Winter's technique a  tight continuity bound for the quantum conditional mutual information in infinite-dimensional tripartite systems under the energy constraint on one subsystem is obtained in \cite[the Appendix]{SE}.

In this paper we specify Fannes' type and Winter's type  continuity bounds for the quantum conditional mutual information (obtained respectively in \cite{CMI}  and \cite{SE}).  Then, by using the Leung-Smith telescopic trick from \cite{L&S} tight continuity bounds of  both types for the output quantum conditional mutual information for $n$-tensor power of a channel are obtained.

We analyse continuity properties of the Holevo quantity with respect to two nonequivalent metrics $D_0$ and $D_*$ on the set of discrete ensembles of quantum states.  The metric $D_0$ is a trace norm distance between ensembles considered as ordered collections of states, the metric $D_*$ is a factorization of $D_0$ obtained by identification of all ensembles corresponding to the same probability measure on the set of quantum states, which  coincides with the EHS-distance between ensembles introduced by Oreshkov and Calsamiglia in \cite{O&C}.
It follows that $D_*$  is upper bounded by the Kantorovich metric $D_K$ and that $D_*$ generates the weak convergence topology on the set of all ensembles considered as probability measures (so, the metrics $D_*$ and $D_K$ are equivalent in the topological sense).

We show that the Holevo quantity is
continuous on the set of all ensembles of $\,m\,$ states with respect to all the metrics $D_0$, $D_*$ and $D_K$ if
either $\,m\,$ or the dimension of underlying Hilbert space is finite and obtain Fannes' type tight continuity bounds for the Holevo quantity with respect to these metrics in this case.

In general case conditions for local continuity of the Holevo quantity with respect to the metrics $D_0$ and $D_*$ and  their corollaries are found. Winter's type  tight continuity bound for the Holevo quantity under the constraint on the average energy of ensembles is obtained and applied to the system  of quantum oscillators. The case of generalized (continuous) ensembles (with the Kantorovich metric $D_K$ in the role of a distance) is considered separately.

The above results are used  to obtain tight and close-to-tight continuity bounds for basic capacities of channels with finite-dimensional output (significantly refining the Leung-Smith continuity bounds from \cite{L&S}). In \cite{SCT} these results are used to prove uniform continuity of the entanglement-assisted and unassisted classical capacities of infinite-dimensional energy-constrained channels (as functions of a channel) with respect to the strong convergence topology (which is substantially weaker than the diamond-norm topology).

\section{Preliminaries}

Let $\mathcal{H}$ be a finite-dimensional or separable infinite-dimensional Hilbert space,
$\mathfrak{B}(\mathcal{H})$ the algebra of all bounded operators with the operator norm $\|\cdot\|$ and $\mathfrak{T}( \mathcal{H})$ the
Banach space of all trace-class
operators in $\mathcal{H}$  with the trace norm $\|\!\cdot\!\|_1$. Let
$\mathfrak{S}(\mathcal{H})$ be  the set of quantum states (positive operators
in $\mathfrak{T}(\mathcal{H})$ with unit trace) \cite{H-SCI,N&Ch,Wilde}.

Denote by $I_{\mathcal{H}}$ the unit operator in a Hilbert space
$\mathcal{H}$ and by $\id_{\mathcal{\H}}$ the identity
transformation of the Banach space $\mathfrak{T}(\mathcal{H})$.\smallskip

A finite or
countable collection $\{\rho_{i}\}$ of states
with a probability distribution $\{p_{i}\}$ is conventionally called
\textit{(discrete) ensemble} and denoted $\{p_{i},\rho_{i}\}$. The state
$\bar{\rho}\doteq\sum_{i}p_{i}\rho_{i}$ is called \emph{average state} of this  ensemble. \smallskip

If quantum systems $A$ and $B$ are described by Hilbert spaces  $\H_A$ and $\H_B$ then the bipartite system $AB$ is described by the tensor product of these spaces, i.e. $\H_{AB}\doteq\H_A\otimes\H_B$. A state in $\S(\H_{AB})$ is denoted $\rho_{AB}$, its marginal states $\Tr_{\H_B}\rho_{AB}$ and $\Tr_{\H_A}\rho_{AB}$ are denoted respectively $\rho_{A}$ and $\rho_{B}$. In this paper a special role is plaid by so called \emph{$qc$-states} having the form
\begin{equation}\label{qcs}
\rho_{AB}=\sum_{i=1}^m p_i\rho_i\otimes |i\rangle\langle i|,
\end{equation}
where $\{p_i,\rho_i\}_{i=1}^m$ is an ensemble of $\,m\leq+\infty\,$ quantum states in $\,\S(\H_A)$ and $\,\{|i\rangle\}_{i=1}^m$ is an orthonormal basis in $\H_B$.\smallskip

The \emph{von Neumann entropy} $H(\rho)=\mathrm{Tr}\eta(\rho)$ of a
state $\rho\in\mathfrak{S}(\mathcal{H})$, where $\eta(x)=-x\log x$,
is a concave nonnegative lower semicontinuous function on $\mathfrak{S}(\mathcal{H})$, it is continuous if and only if $\,\dim\H<+\infty$ \cite{L-2,W}.
The concavity of the von Neumann entropy is supplemented by the
inequality
\begin{equation}\label{w-k-ineq}
H\!\left(p\rho+(1-p)\sigma\right)\leq p
H(\rho)+(1-p)H(\sigma)+h_2(p),\quad p\in(0,1),
\end{equation}
where $h_2(p)=\eta(p)+\eta(1-p)$, valid for any
states $\rho$ and $\sigma$ \cite{N&Ch}. \smallskip

Audenaert obtained in \cite{Aud} the sharpest continuity bound for the von Neumann entropy:
\begin{equation}\label{E-CB}
|H(\rho)-H(\sigma)|\leq \varepsilon
\log (d-1)+h_2(\varepsilon)
\end{equation}
for any states $\rho$ and $\sigma$ in $\S(\H)$ such that
$\;\varepsilon=\frac{1}{2}\|\shs\rho-\sigma\|_1\leq1-1/d$, where $d=\dim\H$. This continuity bound is a refinement of the well known Fannes continuity bound \cite{Fannes}.
\smallskip

The \emph{quantum conditional entropy}
\begin{equation}\label{c-e-d}
H(A|B)_{\rho}=H(\rho_{AB})-H(\rho_B)
\end{equation}
of a bipartite state $\rho_{AB}$ with finite marginal entropies is essentially used in analysis of quantum systems \cite{H-SCI,Wilde}. It is concave and satisfies the following inequality
\begin{equation}\label{ce-ac}
H(A|B)_{p\rho+(1-p)\sigma}\leq p H(A|B)_{\rho}+(1-p)H(A|B)_{\sigma}+h_2(p)
\end{equation}
for any  $p\in(0,1)$ and any states $\rho_{AB}$ and $\sigma_{AB}$. Inequality (\ref{ce-ac}) follows from concavity of the entropy and  inequality (\ref{w-k-ineq}).\smallskip

The conditional entropy can be extended the set of all  bipartite states $\rho_{AB}$ with finite $H(\rho_{A})$ by the formula
\begin{equation}\label{ext-ce}
H(A|B)_{\rho}=H(\rho_{A})-I(A\!:\!B)_{\rho}
\end{equation}
preserving all the basic properties of the conditional entropy (including concavity and inequality (\ref{ce-ac})) \cite{Kuz}.
\smallskip

Winter proved in \cite{W-CB} the following refinement of the Alicki-Fannes continuity bound for the conditional entropy (obtained in \cite{A&F}):
\begin{equation}\label{CE-CB}
|H(A|B)_{\rho}-
H(A|B)_{\sigma}|\leq 2\varepsilon
\log d+(1+\varepsilon)h_2\!\left(\frac{\varepsilon}{1+\varepsilon}\right)
\end{equation}
for any states $\,\rho,\sigma\in\S(\H_{AB})$ such that
$\;\varepsilon=\frac{1}{2}\|\shs\rho-\sigma\|_1$, where $d=\dim\H_A$. He showed that this continuity bound is tight\footnote{A continuity bound $\;\displaystyle|f(x)-f(y)|\leq B_a(x,y),\;x,y\in S_a\;$ depending on a parameter $\,a\,$ is called \emph{tight} for large $\,a\,$ if $\;\displaystyle\limsup_{a\rightarrow+\infty}\sup_{x,y\in S_a}\frac{|f(x)-f(y)|}{B_a(x,y)}=1$.} for large $d$  and that the factor $2$ in (\ref{CE-CB}) can be removed if $\rho$ and $\sigma$ are $qc$-states, i.e. states having form (\ref{qcs}).\smallskip

Winter also obtained  tight continuity bounds for the entropy and for the conditional entropy for infinite-dimensional quantum states with bounded energy (see details in \cite{W-CB}).\smallskip

The \emph{quantum relative entropy} for two states $\rho$ and
$\sigma$ in $\mathfrak{S}(\mathcal{H})$ is defined as follows
$$
H(\rho\shs\|\shs\sigma)=\sum_i\langle
i|\,\rho\log\rho-\rho\log\sigma\,|i\rangle,
$$
where $\{|i\rangle\}$ is the orthonormal basis of
eigenvectors of the state $\rho$ and it is assumed that
$H(\rho\shs\|\shs\sigma)=+\infty$ if $\,\mathrm{supp}\rho\shs$ is not
contained in $\shs\mathrm{supp}\shs\sigma$ \cite{L-2}.

Several continuity bounds for the relative entropy are proved by Audenaert and Eisert \cite{A&E,A&E+}.
Tight bound for the relative entropy difference expressed via the entropy difference is obtained by Reeb and Wolf \cite{W&R}.

A \emph{quantum channel} $\,\Phi$ from a system $A$ to a system
$B$ is a completely positive trace preserving linear map
$\mathfrak{T}(\mathcal{H}_A)\rightarrow\mathfrak{T}(\mathcal{H}_B)$,
where $\mathcal{H}_A$ and $\mathcal{H}_B$ are Hilbert spaces
associated with these systems \cite{H-SCI,N&Ch,Wilde}.\smallskip

Denote by $\mathfrak{F}(A,B)$ the set of all quantum channels from a system $A$ to a system
$B$. We will use two metrics  on the set $\mathfrak{F}(A,B)$ induced respectively by the operator norm
$$
\|\Phi\|\doteq \sup_{\rho\in\T(\H_A),\|\rho\|_1=1}\|\Phi(\rho)\|_1
$$
and by the diamond norm
$$
\|\Phi\|_{\diamond}\doteq \sup_{\rho\in\T(\H_{AR}),\|\rho\|_1=1}\|\Phi\otimes \id_R(\rho)\|_1,
$$
of a map $\Phi:\T(\H_A)\rightarrow\T(\H_B)$. The latter  coincides with the norm of complete boundedness of the dual map $\Phi^*:\B(\H_B)\rightarrow\B(\H_A)$ to  $\Phi$ \cite{H-SCI,Wilde}.

\section{Tight continuity bounds for the quantum conditional mutual information (QCMI)}

The \emph{quantum mutual information} of a bipartite state $\,\rho_{AB}\,$ is defined as follows
\begin{equation}\label{mi-d}
I(A\!:\!B)_{\rho}=H(\rho_{AB}\shs\Vert\shs\rho_{A}\otimes
\rho_{B})=H(\rho_{A})+H(\rho_{B})-H(\rho_{AB}),
\end{equation}
where the second expression  is valid if $\,H(\rho_{AB})\,$ is finite \cite{L-mi}.

Basic properties of the relative entropy show that $\,\rho\mapsto
I(A\!:\!B)_{\rho}\,$ is a lower semicontinuous function on the set
$\S(\H_{AB})$ taking values in $[0,+\infty]$. It is well known that
\begin{equation}\label{MI-UB}
I(A\!:\!B)_{\rho}\leq 2\min\left\{H(\rho_A),H(\rho_B)\right\}
\end{equation}
for any state $\rho_{AB}$ and that
\begin{equation}\label{MI-UB+}
I(A\!:\!B)_{\rho}\leq \min\left\{H(\rho_A),H(\rho_B)\right\}
\end{equation}
for any separable state $\rho_{AB}$ \cite{L-mi,MI-B,Wilde}.\smallskip

The \emph{quantum conditional mutual information (QCMI)} of a state $\rho_{ABC}$ of a
tripartite finite-dimensional system  is defined by the formula
\begin{equation}\label{cmi-d}
    I(A\!:\!B|C)_{\rho}\doteq
    H(\rho_{AC})+H(\rho_{BC})-H(\rho_{ABC})-H(\rho_{C}).
\end{equation}
This quantity plays important role in quantum
information theory \cite{D&J,Wilde}, its nonnegativity is a basic result well known as \emph{strong subadditivity
of von Neumann entropy} \cite{Simon}. If system $C$ is trivial then (\ref{cmi-d}) coincides with (\ref{mi-d}).\smallskip

In infinite dimensions formula (\ref{cmi-d}) may contain the uncertainty
$"\infty-\infty"$. Nevertheless the
conditional mutual information can be defined for any state
$\rho_{ABC}$ by one of the equivalent expressions
\begin{equation}\label{cmi-e+}
\!I(A\!:\!B|C)_{\rho}=\sup_{P_A}\left[\shs I(A\!:\!BC)_{Q_A\rho
Q_A}-I(A\!:\!C)_{Q_A\rho Q_A}\shs\right],\; Q_A=P_A\otimes I_{BC},\!
\end{equation}
\begin{equation}\label{cmi-e++}
\!I(A\!:\!B|C)_{\rho}=\sup_{P_B}\left[\shs I(B\!:\!AC)_{Q_B\rho
Q_B}-I(B\!:\!C)_{Q_B\rho Q_B}\shs\right],\; Q_B=P_B\otimes I_{AC},\!
\end{equation}
where the suprema are over all finite rank projectors
$P_A\in\B(\H_A)$ and\break $P_B\in\B(\H_B)$ correspondingly and it is assumed that $I(X\!:\!Y)_{Q_X\rho
Q_X}=c I(X\!:\!Y)_{c^{-1} Q_X\rho
Q_X}$, where $c=\Tr Q_X\rho_{ABC}$ \cite{CMI}.\smallskip

It is shown in \cite[Th.2]{CMI} that expressions (\ref{cmi-e+}) and
(\ref{cmi-e++}) define the same  lower semicontinuous function on the set
$\S(\H_{ABC})$ possessing all basic properties of the conditional mutual
information valid in finite dimensions. In particular, the following relation (chain rule)
\begin{equation}\label{chain}
I(X\!:\!YZ|C)_{\rho}=I(X\!:\!Y|C)_{\rho}+I(X\!:\!Z|YC)_{\rho}
\end{equation}
holds for any state $\rho$ in $\S(\H_{XYZC})$ (with possible values $+\infty$ in both sides).
To prove (\ref{chain}) is suffices to note that it holds if the systems $X,Y,Z$ and $C$ are finite-dimensional and to apply  Corollary 9 in \cite{CMI}.

If one of the marginal
entropies $H(\rho_A)$ and $H(\rho_B)$ is finite
then the conditional mutual information  is given, respectively, by the explicit formula\footnote{The correctness of these formulae follows from upper bound (\ref{MI-UB}).}
\begin{equation}\label{cmi-d+}
I(A\!:\!B|C)_{\rho}=I(A\!:\!BC)_{\rho}-I(A\!:\!C)_{\rho},
\end{equation}
and
\begin{equation}\label{cmi-d++}
I(A\!:\!B|C)_{\rho}=I(B\!:\!AC)_{\rho}-I(B\!:\!C)_{\rho}.
\end{equation}

By applying upper bound (\ref{MI-UB}) to expressions (\ref{cmi-d+}) and (\ref{cmi-d++}) we see that
\begin{equation}\label{CMI-UB}
I(A\!:\!B|C)_{\rho}\leq 2\min\left\{H(\rho_A),H(\rho_B),H(\rho_{AC}),H(\rho_{BC})\right\}
\end{equation}
for any state $\rho_{ABC}$. \smallskip

The quantum conditional mutual information is not concave or convex but the  inequality
\begin{equation}\label{F-c-b}
\begin{array}{cc}
\left|p
I(A\!:\!B|C)_{\rho}+(1-p)I(A\!:\!B|C)_{\sigma}-I(A\!:\!B|C)_{p\rho+(1-p)\sigma}\right|\leq h_2(p)
\end{array}
\end{equation}
holds for $p\in(0,1)$ and any states $\rho_{ABC}$, $\sigma_{ABC}$ with finite $I(A\!:\!B|C)_{\rho}$, $I(A\!:\!B|C)_{\sigma}$.
If $\rho_{ABC}$, $\sigma_{ABC}$ are states with finite marginal entropies then (\ref{F-c-b}) can be easily proved by noting that
\begin{equation}\label{I-rep}
I(A\!:\!B|C)_{\rho}=H(A|C)_{\rho}-H(A|BC)_{\rho},
\end{equation}
and by using  concavity of the conditional entropy and inequality
(\ref{ce-ac}). The validity of inequality (\ref{F-c-b}) for any states $\rho_{ABC}$, $\sigma_{ABC}$  with finite conditional mutual information is proved by approximation (using Theorem 2B in \cite{CMI}).
\smallskip

\subsection{Fannes' type continuity bounds for QCMI.}

Property (\ref{F-c-b}) makes it possible to directly apply Winter's modification of the Alicki-Fannes method (cf.\cite{A&F,W-CB}) to the conditional mutual information.

\smallskip

\begin{property}\label{CMI-FCB} \emph{Let $\,\rho$ and $\,\sigma$ be states in $\S(\H_{ABC})$ and
$\,\tau_{\pm}=\frac{[\shs\rho-\sigma\shs]_\pm}{\Tr[\shs\rho-\sigma\shs]_{\pm}}$.\footnote{$[\omega]_+$ and $[\omega]_-$ are respectively the positive and negative parts of an operator $\omega$.}}
\emph{If $\,I(A\!:\!B|C)_{\rho}$, $\,I(A\!:\!B|C)_{\sigma}$, $\,I(A\!:\!B|C)_{\tau_{+}}$ and $\,I(A\!:\!B|C)_{\tau_{-}}$ are finite then
\begin{equation}\label{CMI-FCB-b}
\!|I(A\!:\!B|C)_{\rho}-I(A\!:\!B|C)_{\sigma}-\varepsilon(I(A\!:\!B|C)_{\tau_{+}}-I(A\!:\!B|C)_{\tau_{-}})|\leq 2g(\varepsilon)
\end{equation}
and hence
\begin{equation}\label{CMI-FCB+}
|I(A\!:\!B|C)_{\rho}-I(A\!:\!B|C)_{\sigma}|\leq D\varepsilon
+2g(\varepsilon),
\end{equation}
where $D\doteq\max\{I(A\!:\!B|C)_{\tau_{-}},I(A\!:\!B|C)_{\tau_{+}}\}$, $\;\varepsilon=\frac{1}{2}\|\shs\rho-\sigma\|_1\,$ and $\,g(\varepsilon )\!\doteq\!(1+\varepsilon)h_2\!\left(\frac{\varepsilon}{1+\varepsilon}\right)=(1+\varepsilon)\log(1+\varepsilon)-\varepsilon\log\varepsilon$.\footnote{Note that the function $\,g(\varepsilon )\,$ is involved in the expression for the entropy of Gaussian states \cite[Ch.12]{H-SCI}.}}\smallskip

\emph{If the states $\rho_X$ and $\sigma_X$, where $X$ is one of the subsystems $A,B,AC,BC$, are supported by some $d$-dimensional subspace of $\H_X$ then (\ref{CMI-FCB+}) holds with $D=2\log d$.}

\smallskip

\emph{If either $\,\rho_{AC}=\sigma_{AC}$ or $\,\rho_{BC}=\sigma_{BC}$ then the factor $\,2$ in the right hand sides of (\ref{CMI-FCB-b}) and (\ref{CMI-FCB+}) can be removed.}
 \end{property}\medskip

\emph{Proof.}  Following \cite{W-CB} note that
\begin{equation}\label{omega-star}
\frac{1}{1+\varepsilon}\,\rho+\frac{\varepsilon}{1+\varepsilon}\,\tau_-=\omega_{*}=
\frac{1}{1+\varepsilon}\,\sigma+\frac{\varepsilon}{1+\varepsilon}\,\tau_+,
\end{equation}
where $\,\omega^{*}=(1+\varepsilon)^{-1}(\rho+[\shs\rho-\sigma\shs]_-)\,$ is a state in $\S(\H_{ABC})$. By applying (\ref{F-c-b}) to the
convex decompositions (\ref{omega-star}) of $\,\omega_*$ we obtain
$$
(1-p)\left[I(A\!:\!B|C)_{\rho}-I(A\!:\!B|C)_{\sigma}\right]\leq p
\left[I(A\!:\!B|C)_{\tau_+}
-I(A\!:\!B|C)_{\tau_-}\right]+2\shs
h_2(p)
$$
and
$$
(1-p)\left[I(A\!:\!B|C)_{\sigma}-I(A\!:\!B|C)_{\rho}\right]\leq p
\left[I(A\!:\!B|C)_{\tau_-}-
I(A\!:\!B|C)_{\tau_+}\right]+2\shs h_2(p).
$$
where $p=\frac{\varepsilon}{1+\varepsilon}$. These inequalities imply (\ref{CMI-FCB-b}). Inequality (\ref{CMI-FCB+}) follows from (\ref{CMI-FCB-b}) by nonnegativity of $I(A\!:\!B|C)$.

The second assertion of the proposition follows from the first one and upper bound (\ref{CMI-UB}), since
the states $[\tau_{\pm}]_X$ are supported by the minimal subspace of $\H_X$ containing the supports of $\rho_X$ and $\sigma_X$.

Assume that $\,\rho_{AC}=\sigma_{AC}$. It follows from (\ref{omega-star}) that $[\tau_{+}]_{AC}=[\tau_{-}]_{AC}$.
Let $\{P^n_A\}$ be a sequence of finite rank projectors in $\H_A$ strongly converging to the unit operator $I_A$ . Consider the states
$$
\omega^n=[\Tr P^n_A \omega_A]^{-1}P^n_A\otimes I_{BC}\, \omega P^n_A\otimes I_{BC},\quad \omega=\rho,\sigma,\tau_{-},\tau_{+},\omega_*.
$$
Then (\ref{omega-star}) implies
\begin{equation}\label{omega-star-n}
\frac{s_n}{s_n+\varepsilon t_n}\,\rho^n+\frac{\varepsilon t_n}{s_n+\varepsilon t_n}\,\tau^n_-=\omega^n_{*}=
\frac{s_n}{s_n+\varepsilon t_n}\,\sigma^n+\frac{\varepsilon t_n}{s_n+\varepsilon t_n}\,\tau^n_+,
\end{equation}
where $\,s_n=\Tr P^n_A\rho_A=\Tr P^n_A\sigma_A\,$ and $\,t_n=\Tr P^n_A[\tau_+]_A=\Tr P^n_A[\tau_-]_A\,$. Since $H(A|C)_{\rho^n}=H(A|C)_{\sigma^n}<+\infty$ and
$H(A|C)_{\tau^n_{+}}=H(A|C)_{\tau^n_{-}}<+\infty$ (where $H(A|C)$ is the extended conditional entropy defined by formula (\ref{ext-ce})), representation (\ref{I-rep}) shows that
\begin{equation}\label{dif-eq}
I(A\!:\!B|C)_{\omega_1}-I(A\!:\!B|C)_{\omega_2}=H(A|BC)_{\omega_2}-H(A|BC)_{\omega_1},
\end{equation}
where $\,(\omega_1,\omega_2)=(\rho^n, \sigma^n),(\tau^n_{+},\tau^n_{-})$.
By applying concavity of the conditional entropy and inequality (\ref{ce-ac}) to the
convex decompositions (\ref{omega-star-n}) of $\,\omega_*^n$ and taking (\ref{dif-eq}) into account we obtain
$$
(1-p_n)\left[I(A\!:\!B|C)_{\rho^n}\!-I(A\!:\!B|C)_{\sigma^n}\right]\leq p_n\!
\left[I(A\!:\!B|C)_{\tau^n_+}\!
-I(A\!:\!B|C)_{\tau^n_-}\right]+\shs
h_2(p_n)
$$
and
$$
(1-p_n)\left[I(A\!:\!B|C)_{\sigma^n}\!-I(A\!:\!B|C)_{\rho^n}\right]\leq p_n\!
\left[I(A\!:\!B|C)_{\tau^n_-}\!-
I(A\!:\!B|C)_{\tau^n_+}\right]+\shs h_2(p_n),
$$
where $p_n=\!\frac{\varepsilon t_n}{s_n+\varepsilon t_n}$. The lower semicontinuity of the function  $\,\omega\rightarrow I(A\!:\!B|C)_{\omega}\,$ and its monotonicity under local operations (Th.2 in \cite{CMI}) make it possible to show that
$$
\lim_{n\rightarrow\infty}I(A\!:\!B|C)_{\omega^n}=I(A\!:\!B|C)_{\omega},\quad \omega=\rho,\sigma,\tau_{-},\tau_{+}.
$$
So, passing to the limit in the above inequalities implies (\ref{CMI-FCB-b}) with $g(\varepsilon)$ instead of $2g(\varepsilon)$. $\square$
\smallskip

Proposition \ref{CMI-FCB} implies the following refinement of Corollary 8 in \cite{CMI}. \smallskip

\begin{corollary}\label{CMI-FCB-c1} \emph{If $\,d\doteq\min\{\dim\H_A,\dim\H_B\}<+\infty\,$ then
\begin{equation}\label{MI-CB+}
|I(A\!:\!B|C)_{\rho}-I(A\!:\!B|C)_{\sigma}|\leq 2\varepsilon
\log d+2g(\varepsilon)
\end{equation}
for any states $\,\rho$ and $\,\sigma$ in $\,\S(\H_{ABC})$, where $\;\varepsilon=\frac{1}{2}\|\shs\rho-\sigma\|_1\,$. Continuity bound (\ref{MI-CB+}) is tight even for trivial $\,C$, i.e. in the case  $\,I(A\!:\!B|C)=I(A\!:\!B)$.}\smallskip

\emph{If either $\,\rho_{AC}=\sigma_{AC}$ or $\,\rho_{BC}=\sigma_{BC}$ then the term $\,2g(\varepsilon)$ in the right hand side of (\ref{MI-CB+}) can be replaced by $\,g(\varepsilon)$.}\smallskip

\end{corollary}

\emph{Proof.} Continuity bound (\ref{MI-CB+}) and its specification directly follow from Proposition \ref{CMI-FCB}.

The tightness of this  bound  with trivial $C$ can be shown  by  using the example from \cite[Remark 3]{W-CB}. Let $\H_A=\H_B=\mathbb{C}^d$, $\rho_{AB}$ be a maximally entangled pure state and $\,\sigma_{AB}=(1-\varepsilon)\rho_{AB}+\frac{\varepsilon}{d^2-1}(I_{AB}-\rho_{AB})$. Then it is easy to see that $\;\frac{1}{2}\|\shs\rho_{AB}-\sigma_{AB}\|_1=\varepsilon\,$ and that
$$
I(A\!:\!B)_{\rho}-I(A\!:\!B)_{\sigma}=H(\sigma_{AB})-H(\rho_{AB})=2\varepsilon\log d+h_2(\varepsilon)+O(\varepsilon/d^2).\; \square
$$

\begin{remark}\label{MI-CB-r}
By using Audenaert's continuity bound (\ref{E-CB}) and  Winter's continuity bound (\ref{CE-CB})  one can obtain via representation (\ref{I-rep}) with trivial $C$ the following continuity bound
\begin{equation*}
|I(A\!:\!B)_{\rho}-I(A\!:\!B)_{\sigma}|\leq\varepsilon
\log (d-1)+ 2\varepsilon
\log d+h_2(\varepsilon)+g(\varepsilon),
\end{equation*}
for the quantum mutual information (for $\varepsilon\leq 1-1/d$). Since $h_2(\varepsilon)< g(\varepsilon)$ for $\varepsilon>0$, this continuity bound is slightly better than
(\ref{MI-CB+}) for $d=2$.
\end{remark}\medskip

Consider the states
\begin{equation}\label{qqcs}
\rho_{ABC}=\sum_{i=1}^m p_i\rho^i_{AC}\otimes |i\rangle\langle i|\quad\textrm{and}\quad\sigma_{ABC}=\sum_{i=1}^m q_i\sigma^i_{AC}\otimes |i\rangle\langle i|,
\end{equation}
where $\{p_i,\rho^i_{AC}\}_{i=1}^m$ and $\{q_i,\sigma^i_{AC}\}_{i=1}^m$ are ensemble of $\,m\leq+\infty\,$ quantum states in $\,\S(\H_{AC})$ and $\,\{|i\rangle\}_{i=1}^m$ is an orthonormal basis in $\H_B$. Such states are called $qqc$-states in \cite{Wilde}. It follows from  upper bound (\ref{MI-UB+}) that
\begin{equation}\label{MI-UB++}
I(A\!:\!B|C)_{\rho}\leq I(AC\!:\!B)_{\rho}\leq \max\left\{H(\rho_{AC}), H(\rho_{B})\right\}
\end{equation}
for any $qqc$-state $\rho_{ABC}$.\smallskip

\begin{corollary}\label{CMI-FCB-c3} \emph{If  $\,\rho_{ABC}$ and $\sigma_{ABC}$ are $qqc$-states (\ref{qqcs}) then
\begin{equation}\label{CMI-CB++}
|I(A\!:\!B|C)_{\rho}-I(A\!:\!B|C)_{\sigma}|\leq \varepsilon
\log d+2g(\varepsilon),
\end{equation}
where $\;d\doteq\min\{\dim\H_{AC},m\}\,$ and $\;\varepsilon=\frac{1}{2}\|\shs\rho-\sigma\|_1$. }\smallskip

\emph{The first term  in  (\ref{CMI-CB++}) can be replaced by
$\,\varepsilon\max\!\left\{S(\{\gamma^{-}_i\}),S(\{\gamma^{+}_i\!\})\right\}$,
where $\gamma^{\pm}_i=(2\varepsilon)^{-1}\left(\|\shs p_i\rho_{AC}^i-q_i\sigma_{AC}^i\|_1\pm(p_i-q_i)\right)$ and $\shs S$ is the Shannon entropy.}\smallskip

\emph{If either $\,\sum_i p_i\rho^i_{AC}=\sum_i q_i\sigma^i_{AC}$ or $\,\{p_i,\rho^i_{C}\}=\{q_i,\sigma^i_{C}\}$ then the factor $2$ in the right hand side of (\ref{CMI-CB++}) can be removed.}
\end{corollary} \medskip

\emph{Proof.} The assertions follow from Proposition \ref{CMI-FCB} and upper bound (\ref{MI-UB++}), since
$$
\tau_{\pm}=\frac{1}{\varepsilon}\sum_{i=1}^m [p_i\rho_{AC}^i-q_i\sigma_{AC}^i]_{\pm}\otimes |i\rangle\langle i|\quad \textrm{and hence} \quad [\tau_{\pm}]_B=\sum_{i=1}^m \gamma^{\pm}_i|i\rangle\langle i|.\;\square
$$

If $\,\rho_{ABC}$ is a $qqc$-state (\ref{qqcs}) then it is easy to show that
$$
I(A\!:\!B|C)_{\rho}=\chi(\{p_i, \rho^i_{AC}\})-\chi(\{p_i, \rho^i_{C}\}),
$$
where $\chi(\{p_i, \rho^i_{X}\})$ is the Holevo quantity of ensemble $\{p_i, \rho^i_{X}\}$. So,
Corollary \ref{CMI-FCB-c3} with trivial $C$ gives continuity bound for the Holevo quantity as a function of ensemble (see  Section 4).
Corollary \ref{CMI-FCB-c3} with nontrivial $C$ can be used in analysis of the loss of the Holevo quantity under action of a quantum channel (called the entropic disturbance in \cite{Wilde+}).

\subsection{Winter's type continuity bound for QCMI.}

If both systems $A$ and $B$ are  infinite-dimensional (and $C$ is arbitrary) then the function $I(A\!:\!B|C)_{\rho}$ (defined in (\ref{cmi-e+}),(\ref{cmi-e++})) is not continuous on $\S(\H_{ABC})$ (only lower semicontinuous) and takes infinite values. Several conditions of local continuity of this function are presented in Corollary 7 in \cite{CMI}, which implies, in particular, that the function $I(A\!:\!B|C)_{\rho}$ is continuous on  subsets of tripartite states $\,\rho_{ABC}$ with bounded energy of
$\rho_{A}$, i.e. subsets of the form
\begin{equation}\label{s-b-e}
\C_{H_A,E}\doteq\{\shs\rho_{ABC}\,|\,\Tr H_A\rho_{A}\leq E\shs\},
\end{equation}
where $H_A$ is the Hamiltonian of system $A$, provided that\footnote{Since condition (\ref{g-cond}) guarantees continuity of the entropy $H(\rho_{A})$ on the set $\C_{H_A,E}$ \cite{W}.}
\begin{equation}\label{g-cond}
\Tr e^{-\lambda H_A}<+\infty\,\textrm{ for all }\,\lambda>0.
\end{equation}
Condition (\ref{g-cond}) implies that $H_A$ has discrete spectrum of finite multiplicity, i.e. $H_A=\sum_{n=0}^{+\infty} E_n|\tau_n\rangle\langle \tau_n|$, where $\{|\tau_n\rangle\}_{n=0}^{+\infty}$ is an orthonormal basis of eigenvectors of $H_A$  corresponding to the nondecreasing sequence $\{E_n\}_{n=0}^{+\infty}$ of eigenvalues (energy levels of $H_A$) such that $\sum_{n=0}^{+\infty} e^{-\lambda E_n}$ is finite for all $\,\lambda>0$.
By condition (\ref{g-cond}) for any $E$ the von Neumann entropy $H(\rho)$ attains its unique maximum under the constraint $\Tr H_A\rho\leq E$ at the Gibbs state $\gamma_A(E)=[\Tr e^{-\lambda(E) H_A}]^{-1}e^{-\lambda(E) H_A}$, where $\lambda(E)$ is the solution of the equation $\Tr H_Ae^{-\lambda H_A}=E\Tr e^{-\lambda H_A}$ \cite{W}. \smallskip

By Proposition 1 in \cite{EC}  condition (\ref{g-cond}) implies that
\begin{equation}\label{F-def}
F_{H_A}(E)\doteq\sup_{\Tr H_A\rho\leq E}H(\rho)=H(\gamma_A(E))=o\shs(E)\quad\textrm{as}\quad E\rightarrow+\infty.
\end{equation}

Recently Winter obtained  tight continuity bounds for the entropy and for the conditional entropy in infinite-dimensional systems under the energy constraints \cite{W-CB}. By using Winter's technique a  tight continuity bound for the quantum conditional mutual information $I(A\!:\!B|C)$  under the energy constraint on the subsystem $A$ is obtained in \cite[the Appendix]{SE}. In the proofs of these continuity bounds it is assumed that
the Hamiltonian $H_A$ satisfies the condition (\ref{g-cond}) and that
\begin{equation}\label{g-cond+}
E_0\doteq \inf_{\|\varphi\|=1}\langle\varphi|H_A|\varphi\rangle=0.
\end{equation}
The latter condition is used implicitly, it makes the above-defined nonnegative function $F_{H_A}(E)$ concave and strictly increasing on $[0,+\infty)$. To remove condition (\ref{g-cond+}) it suffices to note that Winter's arguments remain valid if we replace the function $F_{H_A}(E)=H(\gamma_A(E))$ by any its upper bound $\widehat{F}_{H_A}(E)$ defined on $[0,+\infty)$ possessing the properties
\begin{equation}\label{F-prop-1}
\widehat{F}_{H_A}(E)> 0,\quad \widehat{F}_{H_A}^{\shs\prime}(E)>0,\quad \widehat{F}_{H_A}^{\shs\prime\prime}(E)< 0\quad\textrm{for all }\; E>0
\end{equation}
and
\begin{equation}\label{F-prop-2}
\widehat{F}_{H_A}(E)=o\shs(E)\quad\textrm{as}\quad E\rightarrow+\infty.
\end{equation}
Note that at least one such function $\widehat{F}_{H_A}(E)$ always exists. By Proposition 1 in \cite{EC}  one can use $F_{H_A}(E+E_0)$ in  the role of $\widehat{F}_{H_A}(E)$. But for applications such choice may be not optimal (see below).

By using this observation and Corollary \ref{CMI-FCB-c1} one can obtain the following "optimized" continuity bound for QCMI.\smallskip\pagebreak

\begin{property}\label{c-b-mi} \emph{Let $H_A$ be the Hamiltonian of system $A$ satisfying conditions (\ref{g-cond}) and $\widehat{F}_{H_A}(E)$ any upper bound for the function $F_{H_A}(E)$  (defined in (\ref{F-def})) with properties  (\ref{F-prop-1}) and (\ref{F-prop-2}), in particular, $\,\widehat{F}_{H_A}(E)=F_{H_A}(E+E_0)$.}\smallskip

\emph{Let $\,\rho\,$ and $\,\sigma\,$ be  states in $\,\S(\H_{ABC})$ such that $\,\Tr H_A\rho_A ,\Tr H_A\sigma_A\leq E$ and  $\frac{1}{2}\|\rho-\sigma\|_1\leq\varepsilon$. Then
\begin{equation}\label{CMI-CB}
\begin{array}{rl}
\left|I(A\!:\!B|C)_{\rho}-I(A\!:\!B|C)_{\sigma}\right|\,\leq &\!\!\!2\varepsilon(2t+r_{\!\varepsilon}(t))\widehat{F}_{H_A}\!\!\left(\frac{E}{\varepsilon t}\right)\\\\
\,+& \!\!\!2g(\varepsilon r_{\!\varepsilon}(t))+4h_2(\varepsilon t),
\end{array}
\end{equation}
for any $\,t\in(0,\frac{1}{2\varepsilon}]$, where $r_{\!\varepsilon}(t)=(1+t/2)/(1-\varepsilon t)\leq r_{\!1}(t)=(1+t/2)/(1-t)$.}
\end{property}\medskip

\begin{remark}\label{c-b-mi-r++}
Continuity bound (\ref{CMI-CB}) can be rewritten in the equivalent Winter form (cf.\cite{W-CB}):
$$
\left|I(A\!:\!B|C)_{\rho}-I(A\!:\!B|C)_{\sigma}\right|\leq2(2\delta_{\varepsilon'}+\varepsilon')\widehat{F}_{H_A}\!(E/\delta_{\varepsilon'})
+2g(\varepsilon')+4h_2(\delta_{\varepsilon'}),
$$
for any $\,\varepsilon'\in(\varepsilon,1]$ such that $\delta_{\varepsilon'}=(\varepsilon'-\varepsilon)/(\varepsilon'+1/2)\leq1/2$, by the change of variables $\varepsilon'=\varepsilon r_{\!\varepsilon}(t)$. The form (\ref{CMI-CB}) seems  preferable because of its explicit dependence on $\varepsilon$.
\end{remark}\smallskip

\begin{remark}\label{c-b-mi-r+} It is easy to see that the right hand side of (\ref{CMI-CB}) attains minimum at some \emph{optimal} $\,t=t(E, \varepsilon)$. It is this minimum that gives proper upper bound for $\,\left|I(A\!:\!B|C)_{\rho}-I(A\!:\!B|C)_{\sigma}\right|$.
\end{remark}\smallskip

\begin{remark}\label{c-b-mi-r} Condition (\ref{F-prop-2}) implies $\,\displaystyle \lim_{x\rightarrow+0}x\widehat{F}_{H_A}\!(E/x)=0\,$. Hence, Proposition \ref{c-b-mi} shows that the function $\rho_{ABC}\mapsto I(A\!:\!B|C)_{\rho}$ is uniformly continuous on the set $\C_{H_A,E}$ defined in (\ref{s-b-e}) for any $E>E_0$.
\end{remark}\medskip

\emph{Proof.} Following the proofs of Lemmas 16,17 in \cite{W-CB} take any $\delta\in(0,\frac{1}{2}]$ and denote by $P_{\delta}$  the spectral projector of the operator $H_A$ corresponding to the interval $[0, \delta^{-1}E]$. Consider  the states
$$
\rho^{\delta}=\frac{P_{\delta}\otimes
I_{BC}\,\rho P_{\delta}\otimes
I_{BC}}{\Tr P_{\delta}\rho_{A}}\quad \textrm{and} \quad \sigma^{\delta}=\frac{P_{\delta}\otimes
I_{BC}\,\sigma P_{\delta}\otimes
I_{BC}}{\Tr P_{\delta}\sigma_{A}}.
$$
By using the arguments from the proof of Lemma 16 in \cite{W-CB} it is easy to show  that
\begin{equation}\label{est-1}
  H(\omega_A)-[\Tr P_{\delta}\omega_A] H(\omega^{\delta}_A)\leq \delta \widehat{F}_{H_A}(E/\delta)+h_2(\Tr P_{\delta}\omega_A),
\end{equation}
\begin{equation}\label{est-2}
\Tr P_{\delta}\omega_A\geq 1-\delta,
\end{equation}
where $\,\omega=\rho,\sigma$, and that
\begin{equation}\label{est-3}
\log\Tr P_{\delta}\leq \widehat{F}_{H_A}(E/\delta).
\end{equation}
It follows from (\ref{est-2}) and basic properties of the trace norm that
$$
\|\rho-\sigma\|_1\geq\|r\rho^{\delta}-s\sigma^{\delta}\|_1\geq r\|\rho^{\delta}-\sigma^{\delta}\|_1-|r-s|
\geq(1-\delta)\|\rho^{\delta}-\sigma^{\delta}\|_1-\delta
$$
where $r=\Tr P_{\delta}\rho_A$ and $s=\Tr P_{\delta}\sigma_A$. Hence
\begin{equation}\label{est-3+}
\|\rho^{\delta}-\sigma^{\delta}\|_1\leq 2\varepsilon',\quad \textup{where}\quad \varepsilon'\doteq(\varepsilon+\delta/2)/(1-\delta).
\end{equation}

By using (\ref{est-1}) and (\ref{est-2}) it is easy to derive from Lemma 9 in \cite{SE}  that
\begin{equation}\label{est-4}
\left|I(A\!:\!B|C)_{\omega}-I(A\!:\!B|C)_{\omega^{\delta}}\right|\leq 2\delta \widehat{F}_{H_A}(E/\delta)+2h_2(\delta),\quad \omega=\rho,\sigma.
\end{equation}
By using (\ref{est-3}) and (\ref{est-3+}) we obtain from  Corollary \ref{CMI-FCB-c1} that
\begin{equation}\label{est-5}
\begin{array}{c}
\left|I(A\!:\!B|C)_{\rho^{\delta}}-I(A\!:\!B|C)_{\sigma^{\delta}}\right|\leq2\varepsilon'\log\Tr P_{\delta}+2g(\varepsilon')\\\\\leq
2\varepsilon'\widehat{F}_{H_A}(E/\delta)+2g(\varepsilon').
\end{array}
\end{equation}
It follows from (\ref{est-4}) and (\ref{est-5}) that
$$
\begin{array}{c}
\left|I(A\!:\!B|C)_{\rho}-I(A\!:\!B|C)_{\sigma}\right|\leq\left|I(A\!:\!B|C)_{\rho^{\delta}}-I(A\!:\!B|C)_{\sigma^{\delta}}\right|\\\\
\qquad+\left|I(A\!:\!B|C)_{\rho}-I(A\!:\!B|C)_{\rho^{\delta}}\right|+\left|I(A\!:\!B|C)_{\sigma}-I(A\!:\!B|C)_{\sigma^{\delta}}\right|\\\\
\leq(2\varepsilon'+4\delta)\widehat{F}_{H_A}(E/\delta)
+2g(\varepsilon')+4h_2(\delta).
\end{array}
$$
By taking $\delta=\varepsilon t$, where $t\in(0,\frac{1}{2\varepsilon}]$, we obtain the required inequality. $\square$  \smallskip

Assume now that $A$ is the $\,\ell$-mode quantum oscillator (while $B$ and $C$ are arbitrary systems). Then
\begin{equation}\label{osc-H}
H_A=\sum_{i=1}^{\ell}\hbar\shs\omega_i \left (a^{+}_ia_i+\textstyle\frac{1}{2}I_A\right),
\end{equation}
where $\,a_i\,$ and $\,a^{+}_i\,$ are the annihilation and creation operators and $\,\omega_i\,$ is the frequency of the $i$-th oscillator \cite[Ch.12]{H-SCI}. It follows that\vspace{-5pt}
$$
F_{H_A}(E)=\max_{\{E_i\}}\sum_{i=1}^{\ell}g(E_i/\hbar\omega_i-1/2),\quad E\geq E_0\doteq\frac{1}{2}\sum_{i=1}^{\ell}\hbar\omega_i,\vspace{-5pt}
$$
where $\,g(x)=(x+1)\log(x+1)-x\log x\,$ and the maximum is over all $\ell\textup{-}$tuples $E_1$,...,$E_{\ell}$  such that  $\sum_{i=1}^{\ell}E_i=E$ and $E_i\geq\frac{1}{2}\hbar\omega_i$. The exact value of
$F_{H_A}(E)$ can be calculated by applying the Lagrange multiplier method which leads to a transcendental equation. But following \cite{W-CB} one can obtain
tight upper bound for $F_{H_A}(E)$ by using the inequality $\,g(x)\leq\log(x+1)+1\,$  valid for all $\,x>0$. It implies\vspace{-5pt}
$$
F_{H_A}(E)\leq \max_{\sum_{i=1}^{\ell}E_i=E}\sum_{i=1}^{\ell}\log(E_i/\hbar\omega_i+1/2)+\ell.
$$
By calculating this maximum via the Lagrange multiplier method  we obtain
\begin{equation}\label{bF-ub}
F_{H_A}(E)\leq \widehat{F}_{\ell,\omega}(E)\doteq\ell\log \frac{E+E_0}{\ell E_*}+\ell,\quad E_*=\left[\prod_{i=1}^{\ell}\hbar\omega_i\right]^{1/\ell}.\vspace{-5pt}
\end{equation}
Since $\,\log(x+1)+1-g(x)=o(1)\,$ as $\,x\rightarrow+\infty$, upper bound (\ref{bF-ub}) is tight for large $E$. By using this upper bound one can obtain from  Proposition \ref{c-b-mi} the following\smallskip

\begin{corollary}\label{c-b-cmi-c}
\emph{Let $A$ be the $\,\ell$-mode quantum oscillator, $\,\rho\,$ and $\,\sigma\,$  any states in $\,\S(\H_{ABC})$ such that $\,\Tr H_A\rho_A ,\Tr H_A\sigma_A\leq E$ and $\,\frac{1}{2}\|\shs \rho-\sigma\|_1\leq\varepsilon\,$. Then
\begin{equation}\label{c-b-cmi-c+}
\begin{array}{rl}
\left|I(A\!:\!B|C)_{\rho}-I(A\!:\!B|C)_{\sigma}\right|\,\leq &\!\! 2\varepsilon(2t+r_{\!\varepsilon}(t))\!\left[\displaystyle
\widehat{F}_{\ell,\omega}(E)-\ell\log(\varepsilon t)\right]\\\\ \,
+& \!\!2g(\varepsilon r_{\!\varepsilon}(t))+4h_2(\varepsilon t),
\end{array}
\end{equation}
for any $\,t\in(0,\frac{1}{2\varepsilon}]$, where $r_{\!\varepsilon}(t)=(1+t/2)/(1-\varepsilon t)$ and $\widehat{F}_{\ell,\omega}(E)$ is defined in (\ref{bF-ub}). The function $r_{\!\varepsilon}(t)$ in (\ref{c-b-cmi-c+}) can be replaced by $r_{\!1}(t)=(1+t/2)/(1-t)$}.\smallskip

\emph{Continuity bound (\ref{c-b-cmi-c+}) with optimal $\,t$ is  tight for large $E$ even for trivial $C$, i.e. in the case $\,I(A\!:\!B|C)=I(A\!:\!B)$.}
\end{corollary}\medskip

\emph{Proof.} Since
$\widehat{F}_{\ell,\omega}(E/x)\leq \widehat{F}_{\ell,\omega}(E)-\ell\log x$ for any positive $E$ and $x\leq1$,
the main assertion of the corollary directly follows from Proposition \ref{c-b-mi}.

Let $\rho_{AB}$ be a purification of the Gibbs state
$\gamma_A(E)$ and $\sigma_{AB}=(1-\varepsilon)\rho_{AB}+\varepsilon\shs \varsigma_A\otimes\varsigma_B$, where $\varsigma_A$ is a state in $\S(\H_A)$ such that $\Tr H_A\varsigma_A\leq E$ and $\varsigma_B$ is any state in $\S(\H_B)$.
Then inequality (\ref{F-c-b}) implies
$$
I(A\!:\!B)_{\rho}-I(A\!:\!B)_{\sigma}\geq2\varepsilon H(\gamma_A(E))-h_2(\varepsilon)=2\varepsilon F_{H_A}(E)-h_2(\varepsilon).
$$
Since $\,\widehat{F}_{\ell,\omega}(E)-F_{H_A}(E)=o(1)\,$ as $\,E\rightarrow+\infty$, the  tightness of the continuity bound
(\ref{c-b-cmi-c+}) for trivial $C$ follows from Remark \ref{os} below. $\square$

\begin{remark}\label{os}
The parameter $\,t$   can be used to optimize  continuity bound (\ref{c-b-cmi-c+}) for given value of energy $E$. The below Lemma \ref{el} (proved by elementary methods) implies that for large energy $E$  the main term in this continuity bound can be made  not greater than $\,\varepsilon (2\widehat{F}_{\ell,\omega}(E)+o\shs(\widehat{F}_{\ell,\omega}(E)))\,$ by appropriate choice of $\,t$.
\end{remark}\smallskip

\begin{lemma}\label{el} \emph{Let $\,f(t)=2t+(1+t/2)/(1-t)$, $\,b>0\,$ and $\,c\,$ be arbitrary. Then
$$
\min_{t\in(0,\frac{1}{2})}f(t)(x-b\log t+c)\leq x+o(x)\quad\textrm{as}\quad x\rightarrow+\infty.
$$}
\end{lemma}

\subsection{QCMI at the output of n copies of a channel}

The following proposition is a QCMI-analog of Theorem 11  in \cite{L&S} proved by the same telescopic trick. It gives Fannes' type and Winter's type tight continuity bounds for the function $\Phi\mapsto I(B^n\!:\!D|C)_{\Phi^{\otimes n}\otimes\id_{CD}(\rho)}$ for any given $n$ and a state $\rho$ in $\S(\H^{\otimes n}_{A}\otimes\H_{CD})$.\smallskip

\begin{property}\label{omi} \emph{Let $\,\Phi$ and $\,\Psi$ be  channels from  $A$ to $B$,  $C$ and $D$ be any systems. Let  $\shs\rho$ be any state in $\,\S(\H^{\otimes n}_{A}\otimes\H_{CD})$, $n\in\mathbb{N}$, 
$$
d(\Phi,\Psi,\rho)=\textstyle\frac{1}{2}\sup\left\{\|(\Phi-\Psi)\otimes\id_R(\omega)\|_1\,|\;\omega_{A}\in\{\rho_{A_1},...,\rho_{A_n}\}\right\}\leq\textstyle\frac{1}{2}\|\Phi-\Psi\|_\diamond
$$
and
$\;\Delta^n(\Phi,\Psi,\rho)\doteq\left|I(B^n\!:\!D|C)_{\Phi^{\otimes n}\otimes\id_{CD}(\rho)}-I(B^n\!:\!D|C)_{\Psi^{\otimes n}\otimes\id_{CD}(\rho)}\right|$.}\footnote{$\|\cdot\|_\diamond$ is the diamond norm described at the end of Section 2.}\footnote{The use of $\,d(\Phi,\Psi,\rho)\,$ (instead of $\frac{1}{2}\|\Phi-\Psi\|_\diamond$) as a measure of divergence between $\Phi$ and $\Psi$ makes the assertions of Proposition \ref{omi} substantially stronger (see \cite{SCT}).}\smallskip

A) \emph{If $\,d_B\doteq\dim\H_B<+\infty$ and $\,d(\Phi,\Psi,\rho)\leq\varepsilon$ then
\begin{equation}\label{CBn-1}
\Delta^n(\Phi,\Psi,\rho)\leq 2n\varepsilon
\log d_B +ng(\varepsilon).
\end{equation}
Continuity bounds (\ref{CBn-1}) is tight (for any given $n$ and arbitrary system $C$).}\smallskip

B) \emph{If the Hamiltonian $H_B$ of the system $B$ satisfies  condition (\ref{g-cond}),\break  $\Tr H_B\Phi(\rho_{A_k}), \Tr H_B\Psi(\rho_{A_k})\leq E_k$ for $k=\overline{1,n}$ and $\,d(\Phi,\Psi,\rho)\leq\varepsilon$ then
\begin{equation}\label{CBn-2}
\!\!\!\begin{array}{rl}
\Delta^n(\Phi,\Psi,\rho)\!\!& \leq \,2\varepsilon(2t+r_{\!\varepsilon}(t))\sum\limits_{k=1}^n\widehat{F}_{H_B}\!\left(\frac{E_k}{\varepsilon t}\right)
+2ng(\varepsilon r_{\!\varepsilon}(t))+4nh_2(\varepsilon t)\!\!\\\\&\, \leq 2n\varepsilon(2t+r_{\!\varepsilon}(t))\widehat{F}_{H_B}\!\left(\frac{E}{\varepsilon t}\right)
+2ng(\varepsilon r_{\!\varepsilon}(t))+4nh_2(\varepsilon t)
\end{array}
\end{equation}
for any $\,t\in(0,\frac{1}{2\varepsilon}]$, where $\; E=n^{-1}\sum_{k=1}^nE_k$,  $\,\widehat{F}_{H_B}(E)$ is any upper bound for the function $F_{H_B}(E)$ (defined in (\ref{F-def})) with properties  (\ref{F-prop-1}) and (\ref{F-prop-2}), in particular $\,\widehat{F}_{H_B}(E)=F_{H_B}(E+E_0)$, and  $\,r_{\!\varepsilon}(t)=(1+t/2)/(1-\varepsilon t)$.}\smallskip

\emph{If $\,B$ is the $\ell$-mode quantum oscillator then (\ref{CBn-2}) can be rewritten as follows
\begin{equation}\label{CBn-2+}
\begin{array}{c}
\Delta^n(\Phi,\Psi,\rho)\leq 2n\varepsilon(2t+r_{\!\varepsilon}(t))\!\left[\displaystyle
\widehat{F}_{\ell,\omega}(E)-\ell\log(\varepsilon t)\right]\\\\
+\,2ng(\varepsilon r_{\!\varepsilon}(t))+4nh_2(\varepsilon t),
\end{array}
\end{equation}
where $\widehat{F}_{\ell,\omega}(E)$ is defined in (\ref{bF-ub}). Continuity bound (\ref{CBn-2+}) with optimal $\,t$ is tight for large $E$ (for any given $n$ and arbitrary system $C$).}\smallskip
\end{property}\medskip

\emph{Proof.} Following the proof of Theorem 11 in \cite{L&S} introduce the states
$$
\sigma_k=\Phi^{\otimes k}\otimes\Psi^{\otimes (n-k)}\otimes\id_{CD}(\rho),\quad k=0,1,...,n.
$$
Note that $\,H([\sigma_k]_{B_j})<+\infty\,$ for all $\,k,j\,$ in both cases A and B.  We have
\begin{equation}\label{tel}
\!\!\!\!\!\begin{array}{c}
\displaystyle \left|I(B^n\!:\!D|C)_{\sigma_n}\!-I(B^n\!:\!D|C)_{\sigma_0}\right|\displaystyle=
\left|\sum_{k=1}^n I(B^n\!:\!D|C)_{\sigma_k}\!-I(B^n\!:\!D|C)_{\sigma_{k-1}}\right|\\ \leq \displaystyle \sum_{k=1}^n \left|I(B^n\!:\!D|C)_{\sigma_k}\!-I(B^n\!:\!D|C)_{\sigma_{k-1}}\right|.
\end{array}\!\!\!
\end{equation}
By using the chain rule (\ref{chain}) we obtain for each $k$
\begin{equation}\label{tel+}
\!\!\begin{array}{ll}
I(B^n\!:\!D|C)_{\sigma_k}\!-I(B^n\!:\!D|C)_{\sigma_{k-1}}&\!\!\!\!\!= I(B_1...B_{k-1}B_{k+1}...B_n \!:\!D|C)_{\sigma_k}\\\\& \!\!\!\!\!+\,I(B_k\!:\!D|B_1...B_{k-1}B_{k+1}...B_nC)_{\sigma_k}\\\\&\!\!\!\!\!-\,
I(B_1...B_{k-1}B_{k+1}...B_n \!:\!D|C)_{\sigma_{k-1}}\\\\&\!\!\!\!\!-\,I(B_k\!:\!D|B_1...B_{k-1}B_{k+1}...B_nC)_{\sigma_{k-1}}\\\\&\!\!\!\!\!=
I(B_k\!:\!D|B_1...B_{k-1}B_{k+1}...B_nC)_{\sigma_k}\\\\&\!\!\!\!\!-\,
I(B_k\!:\!D|B_1...B_{k-1}B_{k+1}...B_nC)_{\sigma_{k-1}},\!\!\!
\end{array}
\end{equation}
where it was  used that $\Tr_{B_k}\sigma_k=\Tr_{B_k}\sigma_{k-1}$. By upper bound (\ref{CMI-UB}) the finiteness of the entropy of the states $\,[\sigma_k]_{B_1},...,[\sigma_k]_{B_n}$ guarantees finiteness of all the terms in (\ref{tel}) and (\ref{tel+}).

Since
$$
\begin{array}{c}
\|\sigma_k-\sigma_{k-1}\|_1=\left\|\id^{\otimes(k-1)}\otimes(\Phi-\Psi)\otimes\id^{\otimes(n-k)}\left(\Phi^{\otimes(k-1)}
\otimes\id\otimes\Psi^{\otimes(n-k)}(\rho)\right)\right\|_1\\\\
\leq \sup\left\{\|(\Phi-\Psi)\otimes\id_R(\omega)\|_1\,|\;\omega_{A}=\rho_{A_k}\right\}\leq 2d(\Phi,\Psi,\rho)\leq 2\varepsilon
\end{array}
$$
and $\,\Tr_{B_k}\sigma_k=\Tr_{B_k}\sigma_{k-1}$, by applying  Corollary \ref{CMI-FCB-c1}  to the right hand side of (\ref{tel+}) in case A we obtain that the value of
\begin{equation}\label{k-value}
\left|I(B^n\!:\!D|C)_{\sigma_k}-I(B^n\!:\!D|C)_{\sigma_{k-1}}\right|
\end{equation}
is upper bounded by $\,2\varepsilon
\log d_B +g(\varepsilon)$ for any $k$. Similarly, by using Proposition \ref{c-b-mi} in case B we obtain that  for any $k$  the value of (\ref{k-value}) is upper bounded by
$\,2\varepsilon(2t+r_{\!\varepsilon}(t))\widehat{F}_{H_B}\!\left(\frac{E_k}{\varepsilon t}\right)
+2g(\varepsilon r_{\!\varepsilon}(t))+4h_2(\varepsilon t).\,$ Hence  (\ref{CBn-1}) and the first inequality in (\ref{CBn-2}) follow from (\ref{tel}) (since $\Phi^{\otimes n}\otimes\id_{CD}(\rho)=\sigma_n$ and $\Psi^{\otimes n}\otimes\id_{CD}(\rho)=\sigma_0$). The second  inequality in (\ref{CBn-2}) follows from the concavity of the function $\widehat{F}_{H_B}$.

The tightness of continuity bound (\ref{CBn-1}) for trivial $C$ and any given $n$ follows from the tightness of continuity bound (\ref{Q-CB}) for the quantum capacity. It can be directly shown by using the erasure channels $\Phi_{0}$ and $\Phi_{p}$ (see the proof of Proposition \ref{LS-CB+} in Section 5.2) and any maximally entangled pure state $\rho$ in $\S(\H_{AD})$, where $D\cong A$.\smallskip

Continuity bound (\ref{CBn-2+}) is derived from (\ref{CBn-2}) by taking $\,\widehat{F}_{H_B}=\widehat{F}_{\ell,\omega}$ and by noting that $\,\widehat{F}_{\ell,\omega}(E/x)\leq \widehat{F}_{\ell,\omega}(E)-\ell\log x\,$ for any positive $E$ and $x\leq1$.

To show the  tightness of continuity bound (\ref{CBn-2+}) for trivial $C$ and any given $n$ assume that $\Phi$ is the identity channel
from the $\ell$-mode quantum oscillator $A$ to $B=A$ and $\Psi$ is the completely depolarizing channel with the vacuum output state. If $\rho_{AD}$ is any purification of the Gibbs state $\gamma_A(E)$ then
$$
I(B^n\!:\!D^n)_{\Phi^{\otimes n}\otimes\id_{D^n}(\rho^{\otimes n})}-I(B^n\!:\!D^n)_{\Psi^{\otimes n}\otimes\id_{D^n}(\rho^{\otimes n})}=
2nH(\gamma_A(E))=2nF_{H_A}(E).
$$
This shows the  tightness of the continuity bound (\ref{CBn-2+}) for large $E$, since  $\,\widehat{F}_{\ell,\omega}(E)-F_{H_A}(E)=o(1)\,$ as $\,E\rightarrow+\infty$ and  the main term of (\ref{CBn-2+}) can be made not greater than $\,\varepsilon n[2\widehat{F}_{\ell,\omega}(E)+o(\widehat{F}_{\ell,\omega}(E))]\,$ for large $E$ by appropriate choice of $t$ (see Remark \ref{os} in Section 3.2). $\square$ \smallskip

Proposition \ref{omi}A is used in Sections 5.2 to  obtain tight and close-to-tight continuity bounds for quantum and classical capacities of finite-dimensional channels (significantly refining the Leung-Smith continuity bounds),
Proposition \ref{omi}B is used in \cite{SCT}  to prove uniform continuity of the classical capacity of energy-constrained infinite-dimensional quantum channels with respect to the strong (pointwise) convergence topology on the set of all channels with bounded energy amplification factor.

\section{On continuity of the Holevo quantity}

\subsection{Discrete ensembles}

The Holevo quantity of a discrete  ensemble $\{p_i,\rho_i\}_{i=1}^m$ of $\,m\leq+\infty$ quantum states is defined as
$$
\chi\left(\{p_i,\rho_i\}_{i=1}^m\right)\doteq \sum_{i=1}^m p_i H(\rho_i\|\bar{\rho})=H(\bar{\rho})-\sum_{i=1}^m p_i H(\rho_i),\quad \bar{\rho}=\sum_{i=1}^m p_i\rho_i,
$$
where the second formula is valid if $H(\bar{\rho})<+\infty$. This quantity gives the upper bound for classical information obtained by recognizing states of the ensemble by quantum measurements \cite{H-73}. It plays important role in analysis of information properties of quantum systems and channels \cite{H-SCI,N&Ch,Wilde}.\smallskip

Let $\H_A=\H$ and $\,\{|i\rangle\}_{i=1}^m$ be an orthonormal basis in a $m$-dimensional Hilbert space $\H_B$. Then
\begin{equation}\label{chi-rep}
\chi(\{p_i,\rho_i\}_{i=1}^m)=I(A\!:\!B)_{\hat{\rho}},\textrm{ where }\,\hat{\rho}_{AB}=\sum_{i=1}^m p_i\rho_i\otimes |i\rangle\langle i|.
\end{equation}
If $\,H(\bar{\rho})\,$ and $S(\{p_i\}_{i=1}^m)$ are finite (here $S$ is the Shannon entropy) then (\ref{chi-rep}) is directly verified by noting that $H(\hat{\rho}_{A})=H(\bar{\rho})$, $H(\hat{\rho}_{B})=S(\{p_i\}_{i=1}^m)$ and $H(\hat{\rho}_{AB})=\sum_{i=1}^m p_i H(\rho_i)+S(\{p_i\}_{i=1}^m)$. The validity of (\ref{chi-rep}) in general  case can be easily shown by two step approximation using Theorem 1A  in \cite{CMI}.\smallskip

To analyse continuity of the Holevo quantity as a function of ensemble we have to choose a metric on the set of all ensembles.

\subsubsection{Three metrics on the set of discrete  ensembles}

If we consider an ensemble  as an ordered collection of states with the corresponding probability distribution then it is natural to use
the quantity
$$
D_0(\mu,\nu)\doteq\frac{1}{2}\sum_i\|\shs p_i\rho_i-q_i\sigma_i\|_1
$$
as a distance between ensembles $\mu=\{p_i,\rho_i\}$ and $\nu=\{q_i,\sigma_i\}$. Since
$D_0(\mu,\nu)$  coincides (up to the factor $1/2$) with the trace norm of the difference between the corresponding $qc$-states $\sum_{i} p_i\rho_i\otimes |i\rangle\langle i|$
and $\sum_{i} q_i\sigma_i\otimes |i\rangle\langle i|$, $D_0$  is a true metric on the set of all "ordered" ensembles of quantum states.
Since convergence of a sequence of states to a state in the weak operator topology implies convergence of this sequence in the trace norm \cite{D-A}, a sequence $\{\{p^n_i,\rho^n_i\}\}_n$ of ensembles converges to an ensemble $\{p^0_i,\rho^0_i\}$ with respect to the metric $D_0$ if and only if
\begin{equation}\label{en-conv}
\lim_{n\rightarrow\infty}p^n_i=p^0_i\;\, \textrm{ for all }\, i\;\textrm{ and }\;\lim_{n\rightarrow\infty}\rho^n_i=\rho^0_i\; \textrm{ for all }\, i\,\textrm{  s.t. }\,p^0_i\neq0.\!\!
\end{equation}

But from the quantum information point of view (in particular, in analysis of the Holevo quantity) it is reasonable to consider an ensemble of quantum states $\{p_i,\rho_i\}$ as a discrete probability measure $\sum_i p_i\delta(\rho_i)$  on the set $\S(\H)$ (where $\delta(\rho)$ is the Dirac measure concentrating at a state $\rho$) rather then ordered (or disordered) collection of states. It suffices to say that a singleton ensemble consisting of a state $\sigma$ and the ensemble $\{p_i,\rho_i\}$, where $\rho_i=\sigma$ for all $i$, are identical from the information point of view and correspond to the same measure $\delta(\sigma)$.

For any ensemble $\{p_i,\rho_i\}$ denote by $\E(\{p_i,\rho_i\})$ the set  of all countable ensembles corresponding to the measure $\sum_i p_i\delta(\rho_i)$. The set $\E(\{p_i,\rho_i\})$  consists of ensembles obtained from the ensemble $\{p_i,\rho_i\}$ by composition of the following operations:
\begin{itemize}
             \item permutation of any states;
             \item  splitting: $(p_1,\rho_1),(p_2,\rho_2),...\rightarrow\,(p,\rho_1),(p_1-p,\rho_1), (p_2,\rho_2),...,p\in[0,p_1]$;
             \item joining of equal states: $\,(p_1,\rho_1),(p_2,\rho_1),(p_3,\rho_3),...\rightarrow\,(p_1+p_2,\rho_1),(p_3,\rho_3),...$
\end{itemize}
If we want to identify ensembles corresponding to the same probability measure then it is natural to use the factorization of $D_0$, i.e. the quantity  \begin{equation}\label{f-metric}
D_*(\mu,\nu)\doteq \inf_{\mu'\in \E(\mu),\nu'\in \E(\nu)}D_0(\mu',\nu')
\end{equation}
as a measure of divergence between ensembles $\mu$ and $\nu$. \smallskip

Taking into account  the above-described structure of a set $\E(\{p_i,\rho_i\})$ it is easy to show that the quantity $D_*(\mu,\nu)$ coincides with the
EHS\nobreakdash-\hspace{0pt}distance $D_{\mathrm{ehs}}(\mu,\nu)$ defined by Oreshkov and Calsamiglia as the infimum of the trace norm distances between Extended-Hilbert-Space representations of the ensembles $\mu$ and $\nu$ (see details in \cite{O&C}). It is shown in \cite{O&C} that $D_{\mathrm{ehs}}$ is a true metric on the sets of discrete ensembles (considered as probability measures) having operational interpretations and possessing several natural properties (convexity, monotonicity under action of quantum channels and generalized measurements, etc.). Moreover, the EHS\nobreakdash-\hspace{0pt}distance between ensembles $\mu=\{p_i,\rho_i\}$ and $\nu=\{q_i,\sigma_i\}$ can be expressed without reference to an extended Hilbert space as follows
\begin{equation}\label{ens-metric}
D_{\mathrm{ehs}}(\mu,\nu)\doteq\frac{1}{2}\inf_{\{P_{ij}\},\{Q_{ij}\}}\,\sum_{i,j} \|\shs P_{ij}\rho_i-Q_{ij}\sigma_j\|_1,
\end{equation}
where the infimum is over all joint probability distributions $\{P_{ij}\}$ with the left marginal $\{p_i\}$ and $\{Q_{ij}\}$ with the right marginal $\{q_j\}$, i.e. such that $\sum_jP_{ij}=p_i$ for all $i$ and $\sum_iQ_{ij}=q_j$ for all $j$ \cite{O&C}.
\smallskip

\begin{remark}\label{metrics-r}
The coincidence of (\ref{f-metric}) and (\ref{ens-metric}) shows, in particular, that for ensembles $\mu$ and $\nu$ consisting of $m$ and $n$ states correspondingly the infimum in (\ref{f-metric}) is attained at some ensembles $\mu'$ and $\nu'$ consisting of $\,\leq mn\,$ states and that it can be calculated by standard linear programming procedure \cite{O&C}.
\end{remark}\smallskip

Definition (\ref{f-metric}) of the metric $\,D_{*}=D_{\mathrm{ehs}}\,$ seems more natural and intuitively clear from the mathematical point of view.
This metric is adequate for continuity analysis of the Holevo quantity, but difficult to compute in general. It is clear that
\begin{equation}\label{d-ineq}
D_*(\mu,\nu)\leq D_0(\mu,\nu)
\end{equation}
for any ensembles $\mu$ and $\nu$. But in some cases the metrics $\,D_0\,$ and $\,D_*\,$  is close to each other or even coincide. This holds, for example,  if we consider small perturbations  of states or probabilities of a given ensemble.\smallskip

The third useful metric is the Kantorovich distance
\begin{equation}\label{K-D-d}
D_K(\mu,\nu)=\frac{1}{2}\inf_{\{P_{ij}\}}\sum P_{ij}\|\rho_i-\sigma_j\|_1
\end{equation}
between ensembles $\mu=\{p_i,\rho_i\}$ and $\nu=\{q_i,\sigma_i\}$, where the infimum is over all joint probability distributions $\{P_{ij}\}$ with the marginals $\{p_i\}$ and $\{q_i\}$, i.e. such that $\sum_jP_{ij}=p_i$ for all $i$ and $\sum_iP_{ij}=q_j$ for all $j$ \cite{Bog, O&C}. By using the coincidence of (\ref{f-metric}) and (\ref{ens-metric}) it is easy to show  that
\begin{equation}\label{d-ineq+}
  D_*(\mu,\nu)\leq D_K(\mu,\nu)
\end{equation}
for any discrete ensembles $\mu$ and $\nu$ \cite{O&C}.

It is essential that the different metrics $D_*$ and  $D_K$ generates the same topology on the set of discrete ensembles.\smallskip

\begin{property}\label{metrics}  \emph{The metrics $\,D_*=D_{\mathrm{ehs}}\,$ and  $\,D_K$ generates the weak convergence topology on the set of all ensembles (considered as probability measures), i.e. the convergence of a  sequence $\{\{p^n_i,\rho^n_i\}\}_n$  to an ensemble $\{p^0_i,\rho^0_i\}$ with respect to any of these metrics means that
\begin{equation}\label{en-conv+}
\lim_{n\rightarrow\infty}\sum_i p^n_if(\rho^n_i)=\sum_i p^0_if(\rho^0_i)
\end{equation}
for any continuous bounded function $f$ on $\,\S(\H)$.}
\end{property}\smallskip

\emph{Proof.} The fact that the Kantorovich distance $D_K$ generates the weak convergence topology is well known \cite{Bog}.

It is shown in \cite{O&C} that convergence of a  sequence $\{\{p^n_i,\rho^n_i\}\}_n$  to an ensemble $\{p^0_i,\rho^0_i\}$ with respect to the metric $\,D_*=D_{\mathrm{ehs}}\,$ implies (\ref{en-conv+}) for any uniformly continuous bounded function $f$ on $\S(\H)$. By Theorem 6.1 in \cite{Par} this means that
the $\,D_*$-convergence is not weaker than the weak convergence. But inequality (\ref{d-ineq+}) shows that
the $\,D_{*}$-convergence is not stronger than the $\,D_K$-convergence equivalent to the weak convergence. $\square$\smallskip

The weak convergence topology is widely used in the measure theory and its applications \cite{Bil,Bog,Par}. It has different characterizations. In particular, Theorem 6.1 in \cite{Par} shows  that the weak convergence of a  sequence $\{\{p^n_i,\rho^n_i\}\}_n$  to an ensemble $\{p^0_i,\rho^0_i\}$ means that
\begin{equation}\label{en-conv++}
\lim_{n\rightarrow\infty}\sum_{i:\rho^n_i\in\S} p^n_i=\sum_{i:\rho^0_i\in\S} p^0_i
\end{equation}
for any subset $\S$ of $\S(\H)$ such that $\{\rho^0_i\}\cap\partial\S=\emptyset$, where $\partial\S$ is the boundary of $\S$. It is easy to see that this convergence is substantially weaker than convergence (\ref{en-conv}).

One of the main advantages of the Kantorovich distance is the existence of its natural extension to the set of all generalized (continuous) ensembles which generates the weak convergence topology on this set (see Section 4.2).

\smallskip

We will explore continuity of the function $\{p_i,\rho_i\}\mapsto \chi(\{p_i,\rho_i\})$ with respect to the metrics $\,D_0\,$ and $\,D_*=D_{\mathrm{ehs}}$, i.e. with respect to the convergence (\ref{en-conv}) and to the weak convergence (\ref{en-conv+}). Taking (\ref{d-ineq+}) into account and  noting that the metrics $\,D_*$ and $\,D_K$ generates the same (weak convergence) topology on the set of discrete ensembles, all the below results  (in particular, Fannes' type and Winter's type continuity bounds) can be reformulated by using the metric $D_K$ instead of $D_*$.

\subsubsection{The case of global continuity}

The following proposition contains continuity bounds for the Holevo quantity with respect to the metrics $D_0$ and $\,D_*=D_{\mathrm{ehs}}\,$ (denoted $D_*$ in what follows).\smallskip

\begin{property}\label{CHI-CB} \emph{Let $\,\{p_i,\rho_i\}$  and  $\,\{q_i,\sigma_i\}$ be  ensembles of states in $\S(\H)$, $\varepsilon_0\!=\!D_0\!\left(\{p_i,\rho_i\},\{q_i,\sigma_i\}\right)$,  $\varepsilon_*\!=\!D_*\!\left(\{p_i,\rho_i\},\{q_i,\sigma_i\}\right)\,$  and $\,g(x)=(1+x)h_2\!\left(\frac{x}{1+x}\right)$}\smallskip

A) \emph{If $\,d\doteq\dim\H$ is finite then}
\begin{equation}\label{CHI-CB+1}
\left|\chi(\{p_i,\rho_i\})-\chi(\{q_i,\sigma_i\})\right|\leq \varepsilon_*
\log d+2g(\varepsilon_*)\leq \varepsilon_0
\log d+2g(\varepsilon_0).
\end{equation}

B) \emph{If $\,\{p_i,\rho_i\}$  and  $\,\{q_i,\sigma_i\}$ are ensembles consisting of $\,m$ and $n\leq m$ states respectively then}
\begin{equation}\label{CHI-CB+2}
\!\!\!\left|\chi(\{p_i,\rho_i\})-\chi(\{q_i,\sigma_i\})\right|\leq
\min\!\left\{\varepsilon_*\log (mn) +2g(\varepsilon_*),\varepsilon_0\log m +2g(\varepsilon_0)\right\}\!\!
\end{equation}
\emph{The term $\,\log m$ in  (\ref{CHI-CB+2}) can be replaced by
$\,\max\!\left\{S(\{\gamma^{-}_i\}),S(\{\gamma^{+}_i\!\})\right\}$,
where $\gamma^{\pm}_i=(2\varepsilon_0)^{-1}\left(\|\shs p_i\rho_i-q_i\sigma_i\|_1\pm(p_i-q_i)\right)\!, i=1,m,$ $\shs S$ is the Shannon entropy and it is assumed that $\,q_i=0$ for $\,i>n$ (if $\,n<m$).}\smallskip

\emph{If $\;\sum_i p_i\rho_i=\sum_i q_i\sigma_i$  then the terms $\,2g(\varepsilon_*)$ and $\,2g(\varepsilon_0)$ in (\ref{CHI-CB+1}) and in (\ref{CHI-CB+2}) can be replaced, respectively, by $\,g(\varepsilon_*)$ and $\,g(\varepsilon_0)$. If $\;\{p_i\}=\{q_i\}$ then the term $\,2g(\varepsilon_0)$ in (\ref{CHI-CB+1}) and in (\ref{CHI-CB+2}) can be replaced by $\,g(\varepsilon_0)$}.\smallskip

\emph{Both continuity bounds in (\ref{CHI-CB+1}) and both continuity bounds in (\ref{CHI-CB+2}) are tight.}
 \end{property}\medskip

\emph{Proof.} Take any  joint probability distributions $\{P_{ij}\}$ with the left marginal $\{p_i\}$ and $\{Q_{ij}\}$ with the right marginal $\{q_j\}$ and consider the $qc$-states
\begin{equation}\label{scq}
\!\hat{\rho}_{ABC}=\sum_{i,j} P_{ij}\rho_i\otimes |i\rangle\langle i|\otimes |j\rangle\langle j|,\quad
\hat{\sigma}_{ABC}=\sum_{i,j}Q_{ij}\sigma_j\otimes |i\rangle\langle i|\otimes |j\rangle\langle j|,
\end{equation}
where $\{|i\rangle\}_{i=1}^m$ and $\{|j\rangle\}_{j=1}^n$ are orthonormal base of Hilbert spaces $\H_B$ and $\H_C$ correspondingly.
Representation (\ref{chi-rep}) and the invariance of the Holevo quantity under splitting of states of an ensemble imply
\begin{equation}\label{chi-rep+}
\chi(\{p_i,\rho_i\})=I(A\!:\!BC)_{\hat{\rho}}\quad\textrm{and}\quad\chi(\{q_j,\sigma_j\})=I(A\!:\!BC)_{\hat{\sigma}}.
\end{equation}
So, the first inequality in A and the inequality
$$
|\chi(\{p_i,\rho_i\})-\chi(\{q_i,\sigma_i\})|\leq\varepsilon_*\log (mn) +2g(\varepsilon_*)
$$
in B
follow from Corollary \ref{CMI-FCB-c3} with trivial $C$ (since the coincidence of (\ref{f-metric}) and (\ref{ens-metric}) shows that $\,2\varepsilon_*=\inf\|\hat{\rho}-\hat{\sigma}\|_1$, where the infimum is over all states of the form (\ref{scq})). The second inequality in A follows from (\ref{d-ineq}).

The inequality $\,|\chi(\{p_i,\rho_i\})-\chi(\{q_i,\sigma_i\})|\leq\varepsilon_0\log m +2g(\varepsilon_0)\,$ in B and its specification follow from
representation (\ref{chi-rep}) and Corollary \ref{CMI-FCB-c3} with trivial $C$.\smallskip

The assertions concerning the cases $\;\sum_i p_i\rho_i=\sum_i q_i\sigma_i$ and  $\;\{p_i\}=\{q_i\}$ follow from the last assertion of Corollary \ref{CMI-FCB-c3}.
\medskip

Let $\,\{|i\rangle\}_{i=1}^d$ be an orthonormal basis in $\H=\mathbb{C}^d$  and  $\shs\rho_c\doteq I_{\H}/d\,$ the chaotic state in $\S(\H)$. For given $\varepsilon\in(0,1)$ consider the ensembles $\mu=\{p_i,\rho_i\}_{i=1}^d$ and $\nu=\{q_i,\sigma_i\}_{i=1}^d$, where
$\rho_i=|i\rangle\langle i|$, $\sigma_i=(1-\varepsilon)|i\rangle\langle i|+\varepsilon\rho_c$, $p_i=q_i=1/d\,$ for all $i$. Then it is easy to see that
$D_*(\mu,\nu)\leq D_0(\mu,\nu)=\varepsilon(1-1/d)$, while concavity of the entropy implies
$$
\chi(\mu)-\chi(\nu)=\log d - \log d + H(\sigma_i)\geq\varepsilon\log d.
$$
Since $\,\dim\H=m=n=d$, this shows tightness of both continuity bounds in (\ref{CHI-CB+1}) and of the second  continuity bound in (\ref{CHI-CB+2}). This example with $d=3$ also shows that the second terms in (\ref{CHI-CB+1}) can not be less than $\varepsilon\log 3/3\approx 0.37 \shs\varepsilon$.

Modifying the above example consider the ensemble $\mu=\{p_i,\rho_i\}_{i=1}^d$, where
$\rho_i=\varepsilon|i\rangle\langle i|+(1-\varepsilon)\rho_c$ and $p_i=1/d\,$ for all $i$,
and the singleton ensemble $\nu=\{\rho_c\}$. Then it is easy to see that $D_*(\mu,\nu)\leq\varepsilon$, while inequality (\ref{w-k-ineq})  implies
$$
\chi(\mu)-\chi(\nu)=\chi(\mu)\geq\varepsilon\log d-h_2(\varepsilon).
$$
Since $\,\dim\H=mn=d$, this shows the tightness of the first continuity bounds in (\ref{CHI-CB+1}) and in (\ref{CHI-CB+2}).
Since $D_0(\mu,\nu)\geq (d-1)/d\,$ for any $\varepsilon$, this example also shows the difference between the continuity bounds depending on $\,\varepsilon_*=D_*(\mu,\nu)$ and on $\,\varepsilon_0=D_0(\mu,\nu)$. $\square$
\smallskip

Proposition \ref{CHI-CB} and inequality (\ref{d-ineq+}) imply\smallskip

\begin{corollary}\label{CHI-CB-c}
\emph{The function $\,\{p_i,\rho_i\}\mapsto \chi(\{p_i,\rho_i\})$ is uniformly continuous on the sets of all ensembles consisting of $\,m\leq+\infty$  states in $\,\S(\H)$ with respect to any of the metrics $D_0$, $D_*$ and $D_K$ if  either $\,\dim\H$ or $\,m$ is finite.}
\end{corollary}\smallskip

It is easy to see that the function $\,\{p_i,\rho_i\}\mapsto \chi(\{p_i,\rho_i\})$ is not continuous on the set of countable  ensembles of states
in $\S(\H)$ with respect to any of the metrics $D_0$, $D_*$ and $D_K$ if $\,\dim\H=+\infty$.\smallskip

Proposition \ref{CHI-CB} contains estimates of two types: the continuity bounds with the main term $\varepsilon
\log d\shs$ depending only on the dimension $d\shs$ of the underlying Hilbert space $\H$ and the continuity bounds with the main term $\varepsilon\log m$ depending only on the size $m$ of ensembles. Continuity bounds of the last type are sometimes called dimension-independent. Recently Audenaert obtained the following dimension-independent continuity bound for the Holevo quantity in the case $\shs p_i\equiv q_i$ \cite[Th.15]{Aud+}:
$$
\left|\chi(\{p_i,\rho_i\})-\chi(\{p_i,\sigma_i\})\right| \leq t\log(1+(m-1)/t)+\log(1+(m-1)t),
$$
where $\,t=\frac{1}{2}\max_i\|\shs \rho_i-\sigma_i\|_1$ is the maximal distance between corresponding states of ensembles.  Proposition \ref{CHI-CB}B in this case gives
\begin{equation}\label{sp-cb}
\left|\chi(\{p_i,\rho_i\})-\chi(\{p_i,\sigma_i\})\right|\leq\varepsilon\log m+g(\varepsilon),
\end{equation}
where $\,\varepsilon=\frac{1}{2}\sum_{i} p_i\|\shs\rho_i-\sigma_i\|_1$ is the average distance between corresponding states of ensembles. Since $\,\varepsilon\leq t\,$ and $\,g(x)\,$ is an increasing function, we may replace $\,\varepsilon\,$ by $\,t\,$ in (\ref{sp-cb}).\smallskip

The following continuity bound for the Holevo quantity not depending on the size $m$ of ensembles is obtained by Oreshkov and Calsamiglia in \cite{O&C}:
$$
\left|\chi(\{p_i,\rho_i\})-\chi(\{q_i,\sigma_i\})\right| \leq 2\varepsilon_K\log(d-1) + 2h_2(\varepsilon_K), \quad \varepsilon_K\leq (d-1)/d,
$$
where $\,d=\dim\H$ and $\,\varepsilon_K=D_K\!\left(\{p_i,\rho_i\},\{q_i,\sigma_i\}\right)$ is the Kantorovich distance (defined in (\ref{K-D-d})). It follows from (\ref{d-ineq+}) that Proposition \ref{CHI-CB}A gives
more sharp continuity bound for the Holevo quantity for $d>2$.

\subsubsection{General case} If $\,\dim\H=+\infty\,$ then the function $\,\{p_i,\rho_i\}\mapsto \chi(\{p_i,\rho_i\})$ is not continuous on the set of all discrete ensembles of states
in $\S(\H)$ with respect to any of the metrics $D_0$, $D_*$ and $D_K$. Conditions for local continuity of this function are presented in the following proposition.\smallskip

\begin{property}\label{CHI-LC}  A) \emph{If $\,\{\{p^n_i,\rho^n_i\}\}_n$ is a sequence of countable ensembles $D_*$-converging to an ensemble $\{p^0_i,\rho^0_i\}\,$ (i.e. weakly converging in the sense (\ref{en-conv+})) such that $\;\lim\limits_{n\rightarrow\infty}H(\bar{\rho}_n)=H(\bar{\rho}_0)<+\infty$,  where $\bar{\rho}_n=\sum_i p^n_i\rho^n_i$,
then
\begin{equation}\label{chi-conv}
\lim_{n\rightarrow\infty}\chi(\{p^n_i,\rho^n_i\})=\chi(\{p^0_i,\rho^0_i\})<+\infty.
\end{equation}}

B) \emph{If $\,\{\{p^n_i,\rho^n_i\}\}_n$ is a sequence  $D_0$-converging to an ensemble $\{p^0_i,\rho^0_i\}$ (i.e. converging in the sense (\ref{en-conv}))
and
\begin{equation}\label{S-cond}
\lim_{n\rightarrow\infty}S\left(\{p^n_i\}\right)=S\left(\{p^0_i\}\right)<+\infty,
\end{equation}
where $S$ is the Shannon entropy, then (\ref{chi-conv}) holds.}\smallskip
\end{property}\medskip

\begin{remark}\label{CHI-LC-r}
Condition (\ref{S-cond}) does not imply
(\ref{chi-conv}) for a  sequence $\{\{p^n_i,\rho^n_i\}\}_n$ $D_*$-converging to an ensemble $\{p^0_i,\rho^0_i\}$.
The simplest example showing this can be constructed as follows.

Let $\{\{p^n_i\}_i\}_n$ be a sequence of countable probability distributions converging  (in the $\ell_1$-metric) to the
degenerate probability distribution $(1, 0, 0, ... )$ such that there exists $\,\lim_{n\rightarrow\infty}S(\{p^n_i\}_i)=C>0$. Let $\{\rho_i\}$ be a countable collection of mutually orthogonal pure states in a separable Hilbert space $\H$ and $\{p^0_i\}$ a probability distribution such that $S(\{p^0_i\}_i)=C$. Then the sequence of ensembles $\{\{p^n_i, \rho_i\}_i\}_n$ converges to the ensemble $\{p^0_i, \sigma_i\}_i$, where $\sigma_i=\rho_1$ for all $i$, with respect to the metric $D_*$ and condition (\ref{S-cond}) holds. But
$\chi(\{p^n_i, \rho_i\})=S(\{p^n_i\}_i)$ does not converge to $\chi(\{p^0_i, \sigma_i\})=0$.
\end{remark}\medskip

\emph{Proof of Proposition \ref{CHI-LC}.} Assertion A is a partial case of Proposition \ref{CHI-LC-ce} in Section 4.2.

Since the convergence (\ref{en-conv}) implies  the trace norm convergence of the sequence
$\{\hat{\rho}^n_{AB}\}$ to the state $\hat{\rho}^0_{AB}$, where $\hat{\rho}^n_{AB}=\sum_{i} p^n_i\rho^n_i\otimes |i\rangle\langle i|$, assertion B is derived from Theorem 1A in \cite{CMI} by using representation (\ref{chi-rep}). $\square$\smallskip

Proposition \ref{CHI-LC}B implies the following observation which can be interpreted as stability of the Holevo quantity with respect to perturbation of states of a given ensemble.\smallskip

\begin{corollary}\label{CHI-LC-c} \emph{Let  $\,\{p_i\}$ be a probability distribution with finite Shannon entropy $S(\{p_i\})$. Then
\begin{equation}\label{chi-conv+}
\lim_{n\rightarrow\infty}\chi(\{p_i,\rho^n_i\})=\chi(\{p_i,\rho^0_i\})\leq S(\{p_i\})
\end{equation}
for any sequences $\{\rho^n_1\}, \{\rho^n_2\}, \ldots $  converging respectively to states $\rho^0_1, \rho^0_2, \ldots $}
\end{corollary}
\medskip

By Corollary \ref{CHI-LC-c}  the finiteness of $S(\{p_i\})$ guarantees the validity of (\ref{chi-conv+}) even in the case when  the entropy is not continuous for all the sequences $\{\rho^n_1\}, \{\rho^n_2\}, \ldots $, i.e. when  $H(\rho^n_k)\nrightarrow H(\rho^0_k)$ for all $k=1,2,...$ \smallskip

Proposition \ref{CHI-LC}A
shows that for any $E>0$ the Holevo quantity is continuous
on the set of ensembles $\{p_i,\rho_i\}$ with the average energy $\,\Tr H_A\bar{\rho}=\sum_ip_i\Tr H_A\rho_i\,$ not exceeding $E$
provided the Hamiltonian $H_A$  satisfies condition (\ref{g-cond}). The following proposition gives Winter's type continuity bound for the Holevo quantity with respect to the metric $D_*$ under the average energy constraint.\smallskip

\begin{property}\label{c-b-chi}  \emph{Let $H_A$ be the Hamiltonian of system $A$ satisfying conditions (\ref{g-cond}) and $\widehat{F}_{H_A}(E)$ any upper bound for the function $F_{H_A}(E)$  (defined in (\ref{F-def})) with properties  (\ref{F-prop-1}) and (\ref{F-prop-2}), in particular, $\,\widehat{F}_{H_A}(E)=F_{H_A}(E+E_0)$.}\smallskip

\emph{Let $\,\{p_i,\rho_i\}$  and  $\,\{q_i,\sigma_i\}$ be  ensembles of states in $\,\S(\H_A)$ with the average states $\bar{\rho}$ and $\bar{\sigma}$ such that $\Tr H_A\bar{\rho},\Tr H_A\bar{\sigma}\leq E$ and  $D_*\!\left(\{p_i,\rho_i\},\{q_i,\sigma_i\}\right)\leq\varepsilon$. Then
\begin{equation}\label{CHI-CB++}
\!\left|\chi(\{p_i,\rho_i\})-\chi(\{q_i,\sigma_i\})\right|\leq\varepsilon(2t+r_{\!\varepsilon}(t))\widehat{F}_{H_A}\!\!\left(\frac{E}{\varepsilon t}\right)
+2g(\varepsilon r_{\!\varepsilon}(t))+2h_2(\varepsilon t)
\end{equation}
for any $\,t\in(0,\frac{1}{2\varepsilon}]$, where $r_{\!\varepsilon}(t)=(1+t/2)/(1-\varepsilon t)\leq r_{\!1}(t)=(1+t/2)/(1-t)$.}\smallskip

\emph{If $\,A$ is the $\,\ell$-mode quantum oscillator then
\begin{equation}\label{c-b-cmi-c++}
\begin{array}{rl}
\left|\chi(\{p_i,\rho_i\})-\chi(\{q_i,\sigma_i\})\right| \,\leq &\!\!\varepsilon(2t+r_{\!\varepsilon}(t))\left[\displaystyle
\widehat{F}_{\ell,\omega}(E)-\ell\log(\varepsilon t)\right]\\\\
\,+ &\!\! 2g(\varepsilon r_{\!\varepsilon}(t))+2h_2(\varepsilon t)
\end{array}
\end{equation}
for any $\,t\in(0,\frac{1}{2\varepsilon}]$, where $\,\widehat{F}_{\ell,\omega}(E)$ is defined in (\ref{bF-ub}). Continuity bound (\ref{c-b-cmi-c++}) with optimal $\,t$ is  tight for large $E$.}
\end{property}\medskip

\begin{remark}\label{c-b-chi-r}
Condition (\ref{F-prop-2}) implies  $\,\displaystyle \lim_{x\rightarrow+0}x \widehat{F}_{H_A}\!(E/x)=0$. Hence, Proposition \ref{c-b-chi} shows that the Holevo quantity is uniformly continuous with respect to the metric $D_*$ on the set of all ensembles $\{p_i,\rho_i\}$ with  bounded average energy.
\end{remark}\smallskip

\begin{remark}\label{c-b-chi-r++}
It follows from (\ref{d-ineq}) and (\ref{d-ineq+}) that the metric $D_*$ in Proposition \ref{c-b-chi} can be replaced by the easy-computable metric $D_0$ and by the Kantorovich metric $D_K$.
\end{remark}\smallskip

\emph{Proof.} By using representation (\ref{chi-rep+}) it is easy to see that continuity bound (\ref{CHI-CB++}) can be proved by showing that
\begin{equation*}
\left|I(A\!:\!B)_{\rho}-I(A\!:\!B)_{\sigma}\right|\leq\varepsilon(2t+r_{\!\varepsilon}(t))\widehat{F}_{H_A}\!\!\left(\frac{E}{\varepsilon t}\right)
+2g(\varepsilon r_{\!\varepsilon}(t))+2h_2(\varepsilon t)
\end{equation*}
for any $qc$-states $\,\rho\,$ and $\,\sigma\,$ such that $\,\Tr H_A\rho_{A},\Tr H_A\sigma_{A}\leq E$, $\|\rho-\sigma\|_1\leq2\varepsilon$ and $\,t\in(0,\frac{1}{2\varepsilon}]$. This inequality is proved by repeating the arguments from the proof of Proposition \ref{c-b-mi} with trivial $C$ by using the below Lemma \ref{main-l+} instead of Lemma 9 in \cite{SE} and  Corollary \ref{CMI-FCB-c3} with trivial $C$ instead of Corollary \ref{CMI-FCB-c1}.

If $A$ is the $\,\ell$-mode quantum oscillator  then (\ref{c-b-cmi-c++}) follows  from (\ref{CHI-CB++}) with  $\,\widehat{F}_{H_A}=\widehat{F}_{\ell,\omega}$, since
$\widehat{F}_{\ell,\omega}(E/x)\leq \widehat{F}_{\ell,\omega}(E)-\ell\log x$ for any positive $E$ and $x\leq1$.

Let $\{p_i,\rho_i\}$ be any pure state ensemble with the average state $\gamma_A(E)$ and
$\,q_i=p_i,\,\sigma_i=(1-\varepsilon)\rho_i+\varepsilon\gamma_A(E)$ for all $i$. Then
$$
2D_*\!\left(\{p_i,\rho_i\},\{q_i,\sigma_i\}\right)\leq\sum_{i}\|\shs p_i\rho_i-q_i\sigma_i\|_1=\sum_{i}\varepsilon p_i\|\shs\rho_i-\gamma_A(E)\|_1\leq2\varepsilon
$$
while concavity of the entropy implies
\begin{equation}\label{tmp}
\left|\chi(\{p_i,\rho_i\})-\chi(\{q_i,\sigma_i\})\right|\geq \varepsilon H\!\left(\gamma_A(E)\right)=\varepsilon F_{H_A}(E).
\end{equation}
Since $\,\widehat{F}_{\ell,\omega}(E)-F_{H_A}(E)=o(1)\,$ as $\,E\rightarrow+\infty$, the  tightness of the continuity bound
(\ref{c-b-cmi-c++}) follows from Remark \ref{os+} below. $\square$\smallskip

\begin{remark}\label{os+}
Lemma \ref{el} in Section 3.2 implies that for large energy $E$  the main term of the continuity bound (\ref{c-b-cmi-c++}) can be made not greater than $\varepsilon (\widehat{F}_{\ell,\omega}(E)+o(\widehat{F}_{\ell,\omega}(E)))$ by appropriate choice of $\,t$.
\end{remark}
\smallskip

\begin{lemma}\label{main-l+} \emph{Let $\,P_A$ be a projector in
$\B(\H_A)$ and $\rho_{AB}$  a $qc$-state (\ref{qcs}) with finite $H(\rho_{A})$. Then
\begin{equation}\label{d-est}
-(1-\tau) H(\tilde{\rho}_{A})\leq I(A\!:\!B)_{\rho}-I(A\!:\!B)_{\tilde{\rho}}\leq
H(\rho_{A})-\tau H(\tilde{\rho}_{A}),
\end{equation}
where $\,\tau=\Tr P_A\rho_A$ and $\;\tilde{\rho}_{AB}=\tau^{-1}P_A\otimes I_{B}\shs\rho_{AB}P_A\otimes
I_{B}$.}\footnote{For arbitrary state $\rho_{AB}$ double inequality (\ref{d-est}) holds with additional factors $2$ in the left and in the right sides (see Lemma 9 in \cite{SE}).}
\end{lemma}\medskip

\emph{Proof.} Both inequalities in (\ref{d-est}) are easily derived from the inequalities
\begin{equation}\label{d-est+}
0\leq I(A\!:\!B)_{\rho}-\tau I(A\!:\!B)_{\tilde{\rho}}\leq
H(\rho_{A})-\tau H(\tilde{\rho}_{A})
\end{equation}
by using nonnegativity of $I(A\!:\!B)$ and upper bound (\ref{MI-UB+}).

Note that representation (\ref{chi-rep}) remains valid for an ensemble $\{p_i,\rho_i\}$ of any positive trace class operators if we assume
that $H$ and $I(A\!:\!B)$ are homogenuous extensions of the von Neumann entropy and of the quantum mutual information to the cones of all positive trace class operators and that
$\chi\left(\{p_i,\rho_i\}\right)=H(\bar{\rho})-\sum_{i} p_i H(\rho_i)$ in the case  $H(\bar{\rho})<+\infty$.
This shows that the double inequality (\ref{d-est+}) can be rewritten as follows
$$
0\leq\chi(\{p_i,\rho_i\})-\chi(\{p_i,P_A\shs\rho_iP_A\})\leq H(\bar{\rho})-H(P_A\shs\bar{\rho}\shs P_A).
$$
The first of these inequalities is easily derived from monotonicity of the quantum relative entropy and concavity of the function $\eta(x)=-x\log x$. The second one follows from the definition of the Holevo quantity, since $H(\rho_i)\geq H(P_A\shs\rho_iP_A)$ for all $i$ \cite{L-2}. $\square$ \medskip

\subsection{Generalized (continuous) ensembles}

In analysis of infinite-dimensional quantum systems and channels the notion of \textit{generalized (continuous) ensemble} defined as
a Borel probability measure on the set of quantum states naturally appears \cite{H-SCI,H-Sh-2}. We denote by $\mathcal{P}(\mathcal{H})$ the set of all Borel probability measures on $\mathfrak{S}(\mathcal{H})$ equipped with the topology of weak convergence
\cite{Bog,Par}.\footnote{The weak convergence of a sequence $\{\mu_n\}$  to a measure $\mu_0$ means that
$\,\lim_{n\rightarrow\infty}\int f(\rho)\mu_n(d\rho)=\int f(\rho)\mu_0(d\rho)\,$
for any continuous bounded function $f$ on $\,\S(\H)$.}
 The set $\mathcal{P}(\mathcal{H})$ is a complete
separable metric space containing the dense subset $\mathcal{P}_0(\mathcal{H})$ of discrete measures (corresponding to discrete ensembles) \cite{Par}. The average state of a generalized
ensemble $\mu \in \mathcal{P}(\mathcal{H})$ is the barycenter of the measure
$\mu $ defined by the Bochner integral
\begin{equation*}
\bar{\rho}(\mu )=\int\rho \mu (d\rho ).
\end{equation*}
The average energy of $\,\mu\,$ is determined by the formula
$$
E(\mu)\doteq \Tr H\bar{\rho}(\mu)=\int\Tr H\rho\,\mu(d\rho)
$$
where $H$ is the Hamiltonian of the system.

\smallskip

The Holevo quantity of a
generalized ensemble $\mu \in \mathcal{P}(\mathcal{H})$ is defined as
\begin{equation*}
\chi(\mu)=\int H(\rho\shs \|\shs \bar{\rho}(\mu))\mu (d\rho )=H(\bar{\rho}(\mu
))-\int H(\rho)\mu (d\rho ),  
\end{equation*}%
where the second formula is valid under the condition $H(\bar{\rho}(\mu))<+\infty$ \cite{H-Sh-2}.\smallskip

The Kantorovich distance (\ref{K-D-d}) between discrete ensembles is extended to generalized ensembles $\mu$ and $\nu$ as follows
\begin{equation}\label{K-D-c}
D_K(\mu,\nu)=\frac{1}{2}\inf_{\Lambda\in\Pi(\mu,\nu)}\int_{\S(\H)\times\S(\H)}\|\rho-\sigma\|_1\Lambda(d\rho,d\sigma),
\end{equation}
where $\Pi(\mu,\nu)$ is the set of all probability measures on $\S(\H)\times\S(\H)$ with the marginals $\mu$ and $\nu$.
Since $\frac{1}{2}\|\rho-\sigma\|_1\leq 1$ for all $\rho$ and $\sigma$, the Kantorovich distance (\ref{K-D-c}) generates the weak convergence topology on $\mathcal{P}(\mathcal{H})$  \cite{Bog}.\smallskip

Note first that the continuity condition for the Holevo quantity of discrete ensembles in Proposition \ref{CHI-LC}A holds for generalized ensembles. \smallskip

\begin{property}\label{CHI-LC-ce} \emph{If $\,\{\mu_n\}_n$ is a sequence of generalized ensembles weakly converging to an ensemble $\mu_0$ such that
$\;\lim\limits_{n\rightarrow\infty}H(\bar{\rho}(\mu_n))=H(\bar{\rho}(\mu_0))<+\infty$ then
\begin{equation}\label{chi-conv-ce}
\lim_{n\rightarrow\infty}\chi(\mu_n)=\chi(\mu_0)<+\infty.
\end{equation}}
\end{property}

\emph{Proof.} We may assume that $H(\bar{\rho}(\mu_n))<+\infty$ for all $n$. So, we have
$$
\chi(\mu_n)=H(\bar{\rho}(\mu_n))-\int H(\rho)\mu_n(d\rho).
$$
Hence, to prove (\ref{chi-conv-ce}) it suffices to note that the functions $\,\mu\mapsto\chi(\mu)\,$ and $\,\mu\mapsto\int H(\rho)\mu(d\rho)\,$
are lower semicontinuous on $\,\mathcal{P}(\H)$ (see Proposition 1 and the proof of the Theorem in \cite{H-Sh-2}). $\square$ \smallskip

The following proposition presents Fannes' type and Winter's type continuity bounds for the Holevo quantity of generalized ensembles with respect to the Kantorovich distance $D_K$ defined in (\ref{K-D-c}).\smallskip

\begin{property}\label{CHI-CB-ce} \emph{Let $\,\mu$  and  $\,\nu$ be  ensembles in $\P(\H_A)$ and $\,\varepsilon=D_K(\mu,\nu)$.}\smallskip

A) \emph{If $\,d\doteq\dim\H_A$ is finite then
\begin{equation}\label{CHI-CB-ce-F}
\left|\chi(\mu)-\chi(\nu)\right|\leq \varepsilon
\log d+2g(\varepsilon).
\end{equation}
The factor $2$ in (\ref{CHI-CB-ce-F}) can be removed provided that  $\,\bar{\rho}(\mu)=\bar{\rho}(\nu)$.}\smallskip

B) \emph{Let $H_A$ be the Hamiltonian of system $A$ satisfying conditions (\ref{g-cond}) and $\widehat{F}_{H_A}(E)$ any upper bound for the function $F_{H_A}(E)$  (defined in (\ref{F-def})) with properties  (\ref{F-prop-1}) and (\ref{F-prop-2}), in particular, $\,\widehat{F}_{H_A}(E)=F_{H_A}(E+E_0)$.
If $\;\Tr H_A\bar{\rho}(\mu),\Tr H_A\bar{\rho}(\nu)\leq E\,$ then
\begin{equation}\label{CHI-CB-ce-W}
\left|\chi(\mu)-\chi(\nu)\right|\leq\varepsilon(2t+r_{\!\varepsilon}(t))\widehat{F}_{H_A}\!\!\left(\frac{E}{\varepsilon t}\right)
+2g(\varepsilon r_{\!\varepsilon}(t))+2h_2(\varepsilon t)
\end{equation}
for any $\,t\in(0,\frac{1}{2\varepsilon}]$, where $r_{\!\varepsilon}(t)=(1+t/2)/(1-\varepsilon t)$.}
\smallskip

\emph{If $\,A$ is the $\ell$-mode quantum oscillator  then the right hand side of (\ref{CHI-CB-ce-W}) can be rewritten as the right hand side of (\ref{c-b-cmi-c++}).}
 \end{property}\medskip

\begin{remark}\label{c-b-chi-r-ce}
Since condition (\ref{F-prop-2}) implies  $\,\displaystyle \lim_{x\rightarrow+0}x \widehat{F}_{H_A}(E/x)=0$, Proposition \ref{CHI-CB-ce}B shows uniform continuity of the Holevo quantity on any subset of $\P(\H_A)$ consisting of ensembles with bounded average energy.
\end{remark}\smallskip

\emph{Proof.} For arbitrary generalized ensembles $\mu$ and $\nu$ there exist sequences $\{\mu_n\}$ and $\{\nu_n\}$ of discrete ensembles weakly converging respectively to $\mu$ and $\nu$ such that
$$
\lim_{n\rightarrow\infty}\chi(\mu_n)=\chi(\mu),\quad\lim_{n\rightarrow\infty}\chi(\nu_n)=\chi(\nu)
$$
and $\bar{\rho}(\mu_n)=\bar{\rho}(\mu),\;\bar{\rho}(\nu_n)=\bar{\rho}(\nu)$ for all $n$. Such sequences can be obtained by using the construction from the proof of Lemma 1 in \cite{H-Sh-2} and taking into account the lower semicontinuity of the function $\mu\mapsto\chi(\mu)$ \cite[Pr.1]{H-Sh-2}.

Hence assertions A and B are derived, respectively, from Propositions
\ref{CHI-CB}A and \ref{c-b-chi} by noting that the metric $D_*$ in all the continuity bounds can be replaced by $D_K$ (this follows from (\ref{d-ineq+})). $\square$

\section{Applications}

\subsection{Tight continuity bounds for the Holevo capacity and for the entanglement-assisted classical capacity of a quantum channel}

The \emph{Holevo capacity} of a quantum channel
$\Phi:A\rightarrow B$  is
defined as follows
\begin{equation}\label{HC-def}
C_{\chi}(\Phi)=\sup_{\{p_i,\rho_i\}}\chi(\{p_i,\Phi(\rho_i)\}),
\end{equation}
where the supremum is over all
ensembles of input states. This quantity determines the ultimate rate of transmission of classical information trough the channel $\Phi$ with non-entangled input encoding, it is closely related to the classical capacity of a quantum channel (see Section 5.2 below) \cite{H-SCI,N&Ch,Wilde}.\smallskip

The \emph{classical
entanglement-assisted capacity} of a quantum channel determines the ultimate rate of transmission of classical information when an entangled state between the input and the output of a channel is used as an
additional resource (see details in \cite{H-SCI,N&Ch,Wilde}). By the Bennett-Shor-Smolin-Thaplyal
theorem the classical
entanglement-assisted capacity of a finite-dimensional quantum channel
$\Phi:A\rightarrow B$ is given by the expression
\begin{equation}\label{EAC-def}
C_{\mathrm{ea}}(\Phi)=\sup_{\rho \in
\mathfrak{S}(\mathcal{H}_A)}I(\Phi, \rho),
\end{equation}
in which $\shs I(\Phi, \rho)\shs$ is the quantum mutual information of the channel $\Phi$ at a state $\rho$ defined as follows
\begin{equation}\label{mi}
 I(\Phi,\rho)=I(B\!:\!R)_{\Phi\otimes\mathrm{Id}_{R}(\hat{\rho})},
\end{equation}
where $\mathcal{H}_R\cong\mathcal{H}_A$ and $\hat{\rho}$
is a pure state in $\S(\H_{AR})$ such that $\hat{\rho}_{A}=\rho$  \cite{BSST,H-SCI,Wilde}.

In analysis of variations of the capacities $C_{\chi}(\Phi)$ and $C_{\mathrm{ea}}(\Phi)$ as functions of a  channel we will use the operator norm
$\|\cdot\|$ and the diamond norm $\|\cdot\|_{\diamond}$ described at the end of Section 2. \smallskip

Proposition \ref{CHI-CB}A and Corollary \ref{CMI-FCB-c1} imply the following \smallskip

\begin{property}\label{cap-tcb} \emph{Let $\,\Phi$ and $\,\Psi$ be quantum  channels from $A$ to $B$ and $g(\varepsilon)=(1+\varepsilon)h_2\!\left(\frac{\varepsilon}{1+\varepsilon}\right)$. Then
\begin{equation}\label{HC-CB}
|C_{\chi}(\Phi)-C_{\chi}(\Psi)|\leq \varepsilon
\log d_B +g(\varepsilon),
\end{equation}
where $\,\varepsilon=\frac{1}{2}\|\Phi-\Psi\|$ and $\,d_B=\dim\H_B$, and
\begin{equation}\label{EAC-CB}
|C_{\mathrm{ea}}(\Phi)-C_{\mathrm{ea}}(\Psi)|\leq 2\varepsilon
\log d +g(\varepsilon),
\end{equation}
where $\,\varepsilon=\frac{1}{2}\|\Phi-\Psi\|_{\diamond}$ and $\,d=\min\{\dim\H_A,\dim\H_B\}$.} \medskip

\emph{Both continuity bounds (\ref{HC-CB}) and (\ref{EAC-CB}) are tight.}
\end{property}\medskip

\emph{Proof.} For given ensemble $\{p_i,\rho_i\}$ Proposition \ref{CHI-CB}A shows that
$$
|\chi(\{p_i,\Phi(\rho_i)\})-\chi(\{p_i,\Psi(\rho_i)\})|\leq \varepsilon_0
\log d_B +g(\varepsilon_0),
$$
where $\varepsilon_0=\frac{1}{2}\sum_i p_i\|\Phi(\rho_i)-\Psi(\rho_i)\|_1\leq\frac{1}{2}\|\Phi-\Psi\|$. This and (\ref{HC-def}) imply (\ref{HC-CB}).\smallskip

Continuity bound (\ref{EAC-CB}) is derived similarly from Corollary \ref{CMI-FCB-c1} and expression (\ref{EAC-def}), since for any pure state
$\hat{\rho}_{AR}$ in (\ref{mi}) we have
$$
\|\Phi\otimes\mathrm{Id}_{R}(\hat{\rho})-\Psi\otimes\mathrm{Id}_{R}(\hat{\rho})\|_1\leq\|\Phi-\Psi\|_{\diamond}.
$$

To show the tightness of both continuity bounds assume that $\H_A=\H_B=\mathbb{C}^d$, $\Phi$ is the identity channel (i.e. $\Phi=\id_{\mathbb{C}^d}$) and $\Psi_p(\rho)=(1-p)\rho+pd^{-1}I_{\mathbb{C}^d}$  is a depolarizing channel ($p\in[0,1]$).
Since $\,C_{\mathrm{ea}}(\Phi)=2C_{\chi}(\Phi)=2\log d\,$,
$$
C_{\chi}(\Psi_p)=\log d+(1-pc)\log(1-pc)+pc\log(p/d)$$
and
$$
C_{\mathrm{ea}}(\Psi_p)=2\log d+(1-p\tilde{c})\log(1-p\tilde{c})+p\tilde{c}\log(p/d^2),
$$
where $\,c=1-1/d\,$ and $\,\tilde{c}=1-1/d^2\,$ \cite{H-SCI,King,Wilde}, we have
$$
C_{\chi}(\Phi)-C_{\chi}(\Psi_p)=pc\log d+h_2(pc)+pc\log c
$$
and
$$
C_{\mathrm{ea}}(\Phi)-C_{\mathrm{ea}}(\Psi_p)=2p\tilde{c}\log d+h_2(p\tilde{c})+p\tilde{c}\log \tilde{c}.
$$
These relations show  tightness of continuity bound (\ref{HC-CB}) and (\ref{EAC-CB}), since it is easy to see that
$\|\Phi-\Psi_p\|\leq\|\Phi-\Psi_p\|_{\diamond}\leq 2p$.  $\square$

\subsection{Refinement of the Leung-Smith continuity bounds for classical and quantum capacities of a channel}

By the Holevo-Schumacher-Westmoreland theorem  the classical capacity of
a finite-dimensional channel $\Phi:A\rightarrow B$  is given by
the expression
\begin{equation}\label{HSW}
C(\Phi)=\lim_{n\rightarrow +\infty }n^{-1}C_{\chi}(\Phi^{\otimes n}),\vspace{-5pt}
\end{equation}
where $C_{\chi}$ is the Holevo capacity defined in the previous subsection \cite{H-SCI,Wilde}.\smallskip

By the Lloyd-Devetak-Shor  theorem  the quantum capacity of
a finite-dimensional channel $\Phi:A\rightarrow B$  is given by
the expression
\begin{equation}\label{LDS}
Q(\Phi)=\lim_{n\rightarrow +\infty }n^{-1}\bar{Q}(\Phi^{\otimes n}),\vspace{-5pt}
\end{equation}
where $\bar{Q}(\Phi)$ is the maximum of the coherent information $I_c(\Phi,\rho)\doteq H(\Phi(\rho))-H(\widehat{\Phi}(\rho))$ over all  states $\rho\in \S(\H_A)$ ($\widehat{\Phi}$ is a complementary channel to $\Phi$). \smallskip

Leung and Smith obtained in \cite{L&S} the following continuity bounds for classical and quantum capacities of a channel with finite-dimensional output
\begin{equation}\label{LS-CB-C}
|C(\Phi)-C(\Psi)|\leq 16\varepsilon
\log d_B +4h_2\!\left(2\varepsilon\right),
\end{equation}
\begin{equation}\label{LS-CB-Q}
|Q(\Phi)-Q(\Psi)|\leq 16\varepsilon
\log d_B +4h_2\!\left(2\varepsilon\right),
\end{equation}
where $\,\varepsilon=\frac{1}{2}\|\Phi-\Psi\|_{\diamond}$ and $\,d_B=\dim\H_B$.\footnote{It is  assumed that expressions (\ref{HSW}) and (\ref{LDS}) remain valid in the case $\dim\H_A=+\infty$.} By using Winter's tight continuity bound (\ref{CE-CB}) for the conditional entropy (instead of the original Alicki-Fannes continuity bound) in the Leung-Smith proof  one can replace the main terms in  (\ref{LS-CB-C}) and (\ref{LS-CB-Q}) by $4\varepsilon
\log d_B$. By using Proposition  \ref{omi}A one can replace the main terms in  (\ref{LS-CB-C}) and (\ref{LS-CB-Q}) by $2\varepsilon
\log d_B$ (which gives tight continuity bound for the quantum capacity and close-to-tight continuity bound for the classical capacity).\smallskip

\begin{property}\label{LS-CB+} \emph{Let $\,\Phi$ and $\,\Psi$ be  channels from $A$ to $B$. Then
\begin{equation}\label{C-CB}
|C(\Phi)-C(\Psi)|\leq 2\varepsilon
\log d_B +g(\varepsilon),
\end{equation}
\begin{equation}\label{Q-CB}
|Q(\Phi)-Q(\Psi)|\leq 2\varepsilon
\log d_B +g(\varepsilon),
\end{equation}
where $\,\varepsilon=\frac{1}{2}\|\Phi-\Psi\|_{\diamond}$, $\,d_B=\dim\H_B$ and $\,g(\varepsilon)=(1+\varepsilon)h_2\!\left(\frac{\varepsilon}{1+\varepsilon}\right)$.}\medskip

\emph{Continuity bound (\ref{Q-CB}) is tight, continuity bound (\ref{C-CB}) is close-to-tight (up to the factor $2$ in the main term).}
\end{property}\medskip

\emph{Proof.} Since
$$
C_{\chi}(\Phi^{\otimes n})=\sup\chi(\{p_i,\Phi^{\otimes n}(\rho_i)\}),
$$
where the supremum is over all ensembles  $\{p_i,\rho_i\}$ of states in $\S(\H^{\otimes n}_A)$, continuity bound (\ref{C-CB}) is  obtained by using  Lemma 12 in \cite{L&S}, representation (\ref{chi-rep}) and Proposition \ref{omi}A in Section 3.3.

To prove  continuity bound (\ref{Q-CB}) note that the coherent information can be represented as follows
$$
I_c(\Phi,\rho)= I(B\!:\!R)_{\Phi\otimes\id_R}(\hat{\rho})-H(\rho),
$$
where $\hat{\rho}\in\S(\H_{AR})$ is a purification of a state $\rho$. Hence for arbitrary quantum channels $\Phi$ and $\Psi$, arbitrary $n$ and any state $\rho$ in $\S(\H^{\otimes n}_{A})$ we have
$$
I_c(\Phi^{\otimes n},\rho)-I_c(\Psi^{\otimes n},\rho)=I(B^n\!:\!R^n)_{\Phi^{\otimes n}\otimes\id_{R^n}}(\hat{\rho})-I(B^n\!:\!R^n)_{\Psi^{\otimes n}\otimes\id_{R^n}}(\hat{\rho})
$$
where $\hat{\rho}\in\S(\H^{\otimes n}_{AR})$ is a purification of the state $\rho$. This representation, Proposition \ref{omi}A in Section 3.3 and Lemma 12 in \cite{L&S} imply  (\ref{Q-CB}).

The tightness of  continuity bound (\ref{Q-CB}) for the quantum capacity can be shown by using the  erasure channels
\begin{equation*}
\Phi_p(\rho)=\left[\begin{array}{cc}
(1-p)\rho &  0 \\
0 &  p\Tr\rho
\end{array}\right], \quad p\in[0,1].
\end{equation*}
from $d$-dimensional system $A$ to $(d+1)$-dimensional system $B$. It is known that
$\,Q(\Phi_p)=(1-2p)\log d\,$ for $\,p\leq 1/2$ and $Q(\Phi_p)=0$ for $p\geq 1/2$ \cite{H-SCI,Wilde}.
Hence $Q(\Phi_0)-Q(\Phi_p)=2p\log d\,$ for $p\leq 1/2$. By noting that
$\|\Phi_0-\Phi_p\|_{\diamond}\leq 2p$ we see that continuity bound (\ref{Q-CB}) is tight (for large $d$).

The proof of tightness of continuity bound (\ref{HC-CB}) for the Holevo capacity shows that the main term in (\ref{C-CB}) is close to the optimal one up to the factor $2$, since $C(\Psi_p)$ coincides with $C_{\chi}(\Psi_p)$ for the depolarizing channel $\Psi_p$ \cite{King}. $\square$ \medskip

\subsection{Other applications}

The results of this paper concerning infinite-dimensional quantum systems and channels can be applied for quantitative continuity analysis of capacities of energy-constrained infinite-dimensional quantum channels with respect to the strong (pointwise) convergence topology (which is substantially weaker than the diamond-norm topology). In particular, Propositions \ref{omi}B and \ref{c-b-chi} are used in \cite{SCT} to obtain uniform continuity bounds for the entanglement-assisted and unassisted classical capacities of energy-constrained infinite-dimensional quantum channels with respect to the
\emph{energy-constrained diamond seminorms} generating the strong convergence topology on the set of quantum channels.

\bigskip

I am grateful to A.Winter for valuable communication and for several technical tricks essentially used in this work.
I am also grateful to A.S.Holevo and G.G.Amosov for useful discussion.\smallskip

The research is funded by the grant of Russian Science Foundation
(project No 14-21-00162).

\end{document}